\documentclass[aps,prd,onecolumn,showpacs,superscriptaddress,groupedaddress]{revtex4}  % submission
\usepackage{dcolumn}   % needed for some tables
\usepackage{bm}        % for math
\usepackage[]{graphicx}
\usepackage{url}
\usepackage{ifpdf}
\usepackage{mathtools}
\usepackage{hyperref}
\hypersetup{breaklinks,colorlinks,citecolor=blue}

\usepackage{setspace}
\usepackage{amssymb}
\usepackage{amsfonts}
\usepackage{amsmath}
\usepackage{mathtools}

\usepackage{subfigure}
\usepackage{color}

\newcommand{\im}{{\rm i}}
\newcommand{\secref}[1]{Sec.~\ref{#1}}
\newcommand{\secrefs}[2]{Secs.~\ref{#1} and~\ref{#2}}
\newcommand{\equref}[1]{Eq.~(\ref{#1})}
\newcommand{\equrefs}[2]{Eqs.~(\ref{#1}) and~(\ref{#2})}
\newcommand{\equrefss}[3]{Eqs.~(\ref{#1}), ~(\ref{#2}), and~(\ref{#3})}
\newcommand{\appref}[1]{Appendix~\ref{#1}}
\newcommand{\figref}[1]{Fig.~\ref{#1}}
\newcommand{\figrefs}[2]{Figs.~\ref{#1} and~\ref{#2}}
\newcommand{\ie}{\textit{i.e.,~}}
\newcommand{\eg}{\textit{e.g.,~}}
\newcommand{\prim}[1]{{#1}^{\prime}}

\newcommand{\equ}[1]{\begin{equation}#1\end{equation}}
\newcommand{\eqn}[1]{\begin{eqnarray}#1\end{eqnarray}}

\newcommand{\ang}{{\boldsymbol{n}}}
\renewcommand{\d}{{\mathrm{d}}}
\newcommand{\rhoang}{{\boldsymbol{\rho}}}
\newcommand{\negsp}[1]{\hspace*{-#1mm}}
\newcommand{\sphere}{{\mathbb{S}^2}}
\newcommand{\Rplus}{{\mathbb{R}^+}}
\newcommand{\convdir}{\ensuremath{\circledast}}
\newcommand{\convaxisym}{\ensuremath{\odot}}

\begin{document}

\title{3D weak lensing with spin wavelets on the ball}

\author{Boris Leistedt}
  \email{boris.leistedt.11@ucl.ac.uk}
  \affiliation{Department of Physics and Astronomy, University College London, London WC1E 6BT, U.K.}

\author{Jason~D.~McEwen}
  \email{jason.mcewen@ucl.ac.uk}
  \affiliation{Mullard Space Science Laboratory, University College London, Surrey RH5 6NT, U.K.}

\author{Thomas~D.~Kitching}
  \email{t.kitching@ucl.ac.uk}
  \affiliation{Mullard Space Science Laboratory, University College London, Surrey RH5 6NT, U.K.}

\author{Hiranya V. Peiris}
  \email{h.peiris@ucl.ac.uk}
  \affiliation{Department of Physics and Astronomy, University College London, London WC1E 6BT, U.K.}

%==============================================================================
\begin{abstract}
We construct the spin flaglet transform, a wavelet transform to analyze spin signals in three dimensions. Spin flaglets can probe signal content localized simultaneously in space and frequency and, moreover, are separable so that their angular and radial properties can be controlled independently. They are particularly suited to analyzing of cosmological observations such as the weak gravitational lensing of galaxies.  Such observations have a unique 3D geometrical setting since they are natively made on the sky, have spin angular symmetries, and are extended in the radial direction by additional distance or redshift information. Flaglets are constructed in the harmonic space defined by the Fourier-Laguerre transform, previously defined for scalar functions and extended here to signals with spin symmetries. Thanks to various sampling theorems, both the Fourier-Laguerre and flaglet transforms are theoretically exact when applied to bandlimited signals. In other words, in numerical computations the only loss of information is due to the finite representation of floating point numbers. We develop a 3D framework relating the weak lensing power spectrum to covariances of flaglet coefficients. We suggest that the resulting novel flaglet weak lensing estimator offers a powerful alternative to common 2D and 3D approaches to accurately capture cosmological information. While standard weak lensing analyses focus on either real or harmonic space representations (\ie\ correlation functions or Fourier-Bessel power spectra, respectively), a wavelet approach inherits the advantages of both techniques, where both complicated sky coverage and uncertainties associated with the physical modeling of small scales can be handled effectively. Our codes to compute the Fourier-Laguerre and flaglet transforms are made publicly available.
\end{abstract}


\maketitle

%==============================================================================
\section{Introduction}

Weak gravitational lensing by large-scale structure, for example cosmic shear, is an observational probe that has the potential to constrain both the geometry and growth of structure of the Universe. It is a sensitive probe of dark energy physics, extensions to general relativity, and neutrino mass and hierarchy (see Refs.~\cite{Albrecht:2006um, Heavens:2006uk, Weinberg:2012es, Peacock:2006kj, Kitching:2008dp} and references therein for more details).  
The basic measurements of weak gravitational lensing from galaxy surveys are the third-flattening, or third-eccentricity, of galaxy images (colloquially refereed to as `ellipticity'), and galaxy sizes. These contain information about the intrinsic unlensed shape of the galaxies and the additional ellipticity (called `shear'), as well as size changes caused by the weak gravitational lensing effect along the line of sight. This angular information can be supplemented with the redshift of the galaxies (either photometric or spectroscopic), yielding a catalogue of galaxy positions, shapes, sizes and redshifts. These can then be compared to predictions from cosmological models in order to constrain their parameters.

Due to the cosmological principle, \ie assumptions of isotropy and homogeneity, the mean of the cosmic shear averaged over all galaxies in a sufficiently large survey is expected to be zero. However, the power spectrum of the shear (pair correlations of modes in 3D Fourier space) is expected to be nonzero and depends on both the geometry of the Universe and the power spectrum of matter perturbations in the 3D volume within which the galaxy sample lies. 
The 3D power spectrum of the galaxy shear estimates themselves is known as `3D cosmic shear'. There exist various strategies and approximations to efficiently compute the 3D cosmic shear power spectrum, including linking angular and radial scales, and binning in redshift---known as `tomography'---that are more widely used than the full 3D case because of computational ease. 

From a theoretical perspective, 3D cosmic shear has been developed in a series of papers \cite{Heavens:2003jx, Castro:2005bg, Munshi:2010ny, Kitching:2010wa} that introduced the relation between the underlying matter power spectrum, the geometric lensing kernel, and the on-sky measurements of galaxies shear and size changes. Because the gravitational lensing signal is caused by a tidal effect around massive objects, the amount of deflection that a photon experiences is related to the second derivative of the local gravitational potential, integrated along the line of sight. The resulting 3D shear distribution is thus well characterised by a 3D cosmic shear power spectrum defined in Fourier-Bessel space (the harmonic space defined by the eigenfunctions of the spherical Laplacian, see also \cite{Dai:2012bc} for a more generic presentation). While the theoretical side of 3D cosmic shear is well established, the application to data and inference of cosmological parameters are less developed; in fact, the 3D cosmic shear framework has only been applied to data twice. In Ref.~\cite{Kitching:2006mq} a 3D cosmic shear analysis was performed using data observed on a small field of approximately one square degree area of sky, with the aim of measuring $\sigma_8$ (the amplitude of matter fluctuations on $8$ Mpc scales) and $\Omega_{\rm M}$ (the dimensionless matter density) as a proof of concept. In Ref.~\cite{Kitching:2014dtq}, a similar analysis was performed on the CFHTLenS survey \cite{Heymans:2012gg} and used to constrain cosmological parameters including the dark energy equation of state, which is parametrised as $w(z)=w_0+w_a z /(1+z)$. On large scales, $\gtrsim1.5h^{-1}$Mpc, it was found that results were consistent with the measurements of the cosmic microwave background (CMB) anisotropies from the {\it Planck} satellite \cite{Ade:2015xua}. But on small scales, $\lesssim1.5h^{-1}$Mpc, results were found to be inconsistent and favoured less clustering of matter than that predicted by {\it Planck}, with the interpretation that this could be due to baryonic feedback effects or systematics in the data. 

Due to various observational constraints, galaxy surveys cover small, and potentially disconnected, regions of the sky. This leads to the problem that the observed power spectrum is related to that of the full sky through a convolution with the window function of the survey. Furthermore, inhomogeneous observation strategies, and galaxy sample selection can result in complicated three dimensional masks or weight maps, which significantly increase the computational requirements of 3D methods. In fact, even 2D masks will mix angular and radial modes in a nontrivial way \cite{Kitching:2014dtq}. Two methods have been proposed to take such masks into account. The first is a pseudo-$C_\ell$ approach where the effect of the mask on the Fourier-Bessel modes is computed in a mixing matrix. Then, either the theory is forward convolved, or the data is deconvolved with the inverse mixing matrix. This has been applied to data only once in the weak lensing context but is well known from power spectrum measurements of the CMB and galaxy clustering (see, \eg Refs.~\cite{Wandelt:2000av, Hivon:2001jp, Leistedt:2013gfa, Pontzen:2010yw}). Such a process is computationally difficult as the mixing matrices tend to be close to singular -- for weak gravitational lensing surveys the mask is also very complex on small scales due to masking of stars, cosmic rays and other image artifacts. The second approach is through a forward modeling of the data using Bayesian hierarchical modeling \cite{Alsing:2015zca}, where the mask is included in the model and the data given infinite variance in the regions of the mask (similar to the approach used to handle a mask by \cite{McEwen:2013cka}). This approach provides the full posterior probability of model parameters but is computationally expensive. 

Another area of difficulty in the analysis of the cosmic shear power spectrum is the interpretation of scale-dependent features in the power spectrum. For a given power spectrum of matter perturbations the cosmic shear power spectrum is readily computable from theory. However, because of nonlinear structure growth (\eg Ref.~\cite{Mead:2015yca}), baryonic feedback effects (\eg Ref.~\cite{Semboloni:2011fe}), photometric redshift systematics, and uncertainty in modeling neutrino physics (\eg Ref.~\cite{Harnois-Deraps:2014sva}), the modeling of the matter power spectrum on small-scales of less than $\sim 1h^{-1}$Mpc is highly uncertain. Therefore, designing statistics and weighting of the data that mitigate the use of the uncertain small-scale regime is particularly important in extracting parameter estimates robustly.  

In this paper we deal with the full 3D cosmic shear case. However, it should be noted that the majority of theoretical papers on weak gravitational lensing power spectra use cosmic shear `tomography', and all but two papers \cite{Kitching:2006mq, Kitching:2014dtq} that use cosmic shear for cosmological parameter estimation do not use power spectra but instead use real-space correlation function measurements. Using correlation functions from the data is ostensibly a good way of accounting for the mask in the data because the measured correlation function is weakly affected by it. However, the covariance properties of the data are much more complicated (\eg Refs.~\cite{Hamilton:2005dx, Takada:2013wfa, Kayo:2012nm}) and result in the need to handle the survey mask accurately due to `beat coupling'. Furthermore, the kernel through which correlation functions depend on the power spectrum of matter perturbations is different for each real-space angular measurement, for each redshift bin, and can extend to very small-scales (\eg Ref.~\cite{MacCrann:2014wfa}). In this paper we explain how these issues may be alleviated by working in wavelet space. 

Wavelets are basis functions or kernels that are well localized in both real and frequency space. They are used in a number of disciplines to solve challenging data analysis, data compression, feature extraction, and pattern recognition problems. In cosmology, wavelets on the sphere have become an integral part of state-of-the-art algorithms to analyze data from the CMB, in particular to isolate the CMB signal from astrophysical foregrounds, extract its statistical properties, and search for potential anomalies (see, \eg Refs.~\cite{Barreiro:2001ne, Fay:2008wv, Marinucci:2007aj,  McEwen:2007ni, Adam:2015wua, Ade:2015ava, Starck:2005fb, McEwen:2004sv, McEwen:2006my, McEwen:2007rz} and references therein).  As with the CMB, observations of galaxy surveys are made on the celestial sphere. Yet, wavelets on the sphere are not typically used in standard analyzes of galaxy surveys because the cosmological information of interest is in the full 3D distribution of galaxy properties; ignoring the radial information causes a significant loss of sensitivity to the geometry and growth of structure of the Universe. Various wavelet methods have been developed to analyze galaxy survey data recently but none of them are suited to 3D weak lensing observables, or 3D signals with spin angular symmetries in general. As mentioned previously, weak lensing observables such as cosmic shear are spin signals on the sky, extended to three dimensions through additional radial information. Thus, uncertainties and observational complications are separable on the sky and along the line of sight (\eg the ability to distinguish galaxies from stars and to estimate their redshifts). 

A number of 2D and 3D methods, including 3D wavelets  \cite{Lanusse:2011cm, Leistedt:2012zx, Durastanti:2014jya}, have been developed to analyze projected weak lensing observables as well as the polarization of the CMB, which is also a spin signal (\eg Refs.~\cite{Starck:2005ek, Geller:2008ui, Pires:2012yn, Deriaz:2012sd, Leonard:2013hia}). However, these approaches do not natively deal with 3D spherical geometry and spin symmetries simultaneously. Wavelets in 3D spherical coordinates were constructed in Ref.~\cite{Lanusse:2011cm} using B-spline kernels and the Fourier-Bessel basis.  While these are directly connected to the Fourier-Bessel formalism, they can probe only 3D isotropic features in scalar signals. Hence, they do not account for the angular spin symmetries or  the unique 2+1D geometry of weak lensing observables. By contrast, Ref.~\cite{Leistedt:2012zx} introduced `flaglets' to probe the angular and radial scales separately. This approach was based on the Fourier-Laguerre transform, a novel separable 3D harmonic transform with a sampling theorem and an analytical connection to the Fourier-Bessel basis.  More recently, Ref.~\cite{Durastanti:2014jya} proposed a construction of separable 3D needlets based on the Fourier transform in the radial direction. 

In this paper, we extend the framework of Ref.~\cite{Leistedt:2012zx} and present novel Fourier-Laguerre and flaglet transforms supporting spin symmetries and also directional features in the angular direction. We detail sampling theorems and efficient algorithms to compute these transforms exactly, exploiting the recently constructed spin directional wavelet transform on the sphere and sampling theorem on the rotation group, and their corresponding fast algorithms, developed in Refs.~\cite{McEwen:2015s2let, McEwen:2013tpa, McEwen:2015so3}. Thanks to the separability of the radial and angular components, the novel spin Fourier-Laguerre and flaglet transforms natively support the 2+1D spherical geometry of the weak lensing data while accounting for angular spin symmetries. They can also be related to the standard 3D weak lensing framework in the Fourier-Bessel basis thanks to the analytical connection between the Fourier-Laguerre and Fourier-Bessel transforms. Finally, we discuss the advantages of working in flaglet space (instead of the standard correlation functions, Fourier, or Fourier-Bessel approaches) that arise from the dual localization properties of flaglets in pixel and frequency space, for example in dealing with complicated angular masks or small scales where physical modeling is less certain.

This paper is structured as follows. In \secref{sec:weaklensingintro} we give a short review of the 3D weak lensing formalism. We adopt the language of observational cosmology and focus on the observables that can be measured from galaxy survey data and connected to cosmological models. In \secrefs{sec:flag}{sec:flaglet} we step back from this formalism and give a formal mathematical presentation of the new analysis techniques (transforms and operators) for radial, spherical and 3D fields. This leads to the definition of the novel spin Fourier-Laguerre and spin flaglet transforms (in \secrefs{sec:flag}{sec:flaglet}, respectively). In \secref{sec:weaklensingapplication}, we combine the two viewpoints, apply the spin flaglet transform to 3D weak lensing, and study the properties of the shear flaglet coefficients. We present conclusions in \secref{sec:conclusion}. Further technical details on the special functions and useful approximations used in this paper are presented in appendixes. 

%==============================================================================
\section{3D weak lensing}\label{sec:weaklensingintro}

In this section we give a brief introduction to the 3D weak lensing formalism, which has been developed in a series of recent papers \cite{Heavens:2003jx, Castro:2005bg, Munshi:2010ny, Kitching:2010wa}. Here we focus on the basic principles of 3D weak lensing and discuss the lensing potential, observable quantities, and how observables and theory can be compared to constrain cosmological models and parameters.  For further background we refer the reader to the excellent exposition of 3D weak lensing given in Ref.~\cite{Castro:2005bg}. Details of practical aspects and data-related complications are addressed in Ref.~\cite{Kitching:2014dtq} (\eg pseudo-$C_\ell$ methods, real and imaginary covariance structures, E and B-mode decomposition, and weighting due to shape measurement biases). We  touch these issues briefly in this section, and in \secref{sec:weaklensingapplication}, but leave detailed investigations on data and simulations to future work.

%==============================================================================
\subsection{Lensing potential}

The weak gravitational lensing effect is commonly expressed in terms of the lensing potential $\phi$, which depends on the integrated deflection angle along the line of sight, sourced by the local Newtonian potential $\Phi$,  
{\begin{equation}
	\phi(r,\ang)=\frac{2}{c^2}\int_0^r{\rm d}r'  \frac{f_K(r - r')}{f_K(r)f_K(r')} \ \Phi(r',\ang),
\end{equation}}
where $r$ is the comoving distance, $\ang$ the angular position on the sky, and $c$ is the speed of light in a vacuum. 
The geometrical factor reads 
\eqn{
        f_K(r) &=& \ \left\{ \begin{array}{lll}    \sin(r),  &{\rm if}  \  K = 1 \vspace*{2mm}\\
          r, \quad\quad  &{\rm if} \  K = 0  \vspace*{2mm}\\
        \sinh(r), \quad\quad& {\rm if} \  K = -1 \end{array}\right.,
}
for cosmologies with positive, flat and negative global curvatures, with curvature $K=1$, $0$, and $-1$, respectively. This expression assumes the linear regime and also the Born approximation, \ie that the path of photons is unperturbed by lenses.

The Newtonian potential $\Phi$ can be related to perturbations in the underlying matter density $\delta$ via Poisson's equation,
\equ{
	\nabla^2 \Phi(r,\ang) = \frac{3 \Omega_{\rm M} H_0^2}{2a(r)} \delta(r,\ang), \label{eq:poisson}
}
where $\Omega_{\rm M}$ is the dimensionless matter density, $H_0$ is the current value of the Hubble parameter, and $a(r)$ is the dimensionless scale factor.  The 3D gradient is defined relative to comoving coordinates. This relation can be expressed conveniently in terms of the Fourier-Bessel basis since the Fourier-Bessel basis functions are eigenfunctions of the Laplacian operator $\nabla^2$, with eigenvalues $-k^2$.  Consequently, the harmonic representation of \equref{eq:poisson} reads
\equ{
	\Phi_{\ell m}(k;r) = - \frac{3 \Omega_{\rm M} H_0^2}{2k^2a(r)}\delta_{\ell m}(k;r), \label{eq:newton}
}
where $\ell$ and $m$ label angular harmonic modes and $k$  labels the radial wavenumber (in units of $h^{-1}$ Mpc). The Fourier-Bessel transform is defined explicitly in \secref{sec:fourier-bessel} by \equrefs{fourierbesselanalysis}{fourierbesselsynthesis}, and discussed extensively in the same section. The dependency on $r$ must be retained in \equref{eq:newton} to account for the time evolution of the field: we apply the Fourier-Bessel transform to the homogenous field existing everywhere at the cosmological time corresponding to comoving distance $r$. 

{The harmonic representation of the lensing potential then reads
\eqn{	
	\phi_{\ell m}(k) = \frac{4}{\pi c^2}  \int_\Rplus  {\rm d}r r^2 j_\ell(kr) \int_0^r {\rm d}r^\prime   \frac{f_K(r - r')}{f_K(r)f_K(r')} \ \int_\Rplus {\rm d}k^\prime k^{\prime 2} j_\ell(k^\prime r^\prime) \Phi_{\ell m}(k^\prime; r), \label{eq:phi}
}}
where $j_\ell$ are the spherical Bessel functions and $\Rplus$ denotes the positive real half-line, \ie $[0,\infty)$.  The difference between \equref{eq:phi} and the expression shown in Ref.~\cite{Castro:2005bg} is due only to the different conventions used for the spherical Bessel transform, where we adopt the convention of Refs.~\cite{Heavens:2003jx,Leistedt:2012zx} (see \appref{app:spinsha} and \secref{sec:fourier-bessel}).

%==============================================================================
\subsection{Observables}

Weak lensing generates distortions in the observations of a background field, which may be characterised by spin quantities of several orders (see, \eg Ref.~\cite{Bacon:2008zj}), with spin number $s$. In the weak lensing regime (\ie away from the critical curve of lensing masses, where there are not multiple images of sources) four distortions can be produced \cite{Bacon:2008zj}: the size magnification ${}_0 \kappa$ \cite{Heavens:2013gol}; the shear ${}_2\gamma$; and the flexion, the combined effect of the first-flexion $_1\mathcal{F}$ (a centroid shift) and third-flexion $_3\mathcal{G}$ (a trefoil distortion). We introduce the generic notation ${}_s\chi$ to denote these spin quantities, which are related to the lensing potential $\phi$ by (see, \eg Ref.~\cite{Castro:2005bg})
{\eqn{
        {}_s\chi(\ang,r) &=& \ \left\{ \begin{array}{lll}    {}_0\kappa(\ang,r) & = \ \frac{1}{4}(\eth\bar{\eth}+\bar{\eth}\eth)\phi(\ang,r),  &{\rm if}  \  s = 0 \label{chidef} \vspace*{2mm}\\
          {}_1\mathcal{F}(\ang,r) & =\ \frac{-1}{6}(\bar{\eth}\eth\eth + \eth\bar{\eth}\eth+\eth\eth\bar{\eth})\phi(\ang,r), \quad\quad  &{\rm if} \  s = 1  \vspace*{2mm}\\
        {}_2\gamma(\ang,r) & =\ \frac{1}{2} (\eth\eth)\phi(\ang,r), \quad\quad& {\rm if} \  s = 2  \vspace*{2mm}\\
        {}_3\mathcal{G}(\ang,r)  & =\ \frac{-1}{2} (\eth\eth\eth) \phi(\ang,r), \quad\quad& {\rm if} \  s = 3 \end{array}\right.,
}}
where $\eth$ and $\bar{\eth}$ are spin raising and spin lowering operators \cite{Goldberg:1966uu, Newman:1966ub, Zaldarriaga:1996xe, Kamionkowski:1996ks}. Further details about these operators and spin spherical harmonics in general are given in \secref{sec:flag} and \appref{app:spinsha}. 
We now focus on the spin-2 quantity of shear ${}_2\gamma$, which is of considerable interest since among the spin quantities of ${}_s \chi$ it has the highest signal-to-noise ratio in observational data (see, \eg Refs.~\cite{Bartelmann:1999yn, Kilbinger:2014cea}). We revisit the other spin quantities of ${}_s\chi$ in \secref{flexion} (where we present their flaglet covariances).

The shear field ${}_{\pm 2}\gamma$ is typically decomposed into its real and imaginary parts, ${}_{\pm 2}\gamma=\gamma_1 \pm \im\gamma_2$, and is related to the lensing potential by \cite{Castro:2005bg}
\eqn{
	{}_2 \gamma(\ang, r) = \gamma_1(\ang,r) +\im\gamma_2(\ang,r) &=& \eth^2\bigl(\phi^E(\ang,r) + \im \phi^B(\ang,r)\bigr)/2,\label{eq:shearp2}\\
	{}_2 \gamma^*(\ang, r) = {}_{-2} \gamma(\ang, r) = \gamma_1(\ang,r) -\im\gamma_2(\ang,r) &=&  \bar{\eth}^2\bigl(\phi^E(\ang,r) - \im \phi^B(\ang,r)\bigr)/2,\label{eq:shearm2}
}
where $*$ denotes complex conjugation and we have represented the lensing potential by its parity even and odd components given by the E- and B-mode signals, respectively, denoted by superscripts $E$ and $B$.  Alternatively, the E- and B-mode signals may be related to the shear by
\eqn{
	\tilde{\phi}^E(\ang,r)  
	= \bigl ( \bar{\eth}^2 {}_2 \gamma(\ang, r) + 
	{\eth}^2 {}_{-2} \gamma(\ang, r) \bigr ), \label{eq:emodephi}\\
	\tilde{\phi}^B(\ang,r) ,
	= -{\rm i} \bigl ( \bar{\eth}^2 {}_2 \gamma(\ang, r) -
	{\eth}^2 {}_{-2} \gamma(\ang, r) \bigr ),\label{eq:bmodephi}
}
where $\tilde{\phi}^E$ and $\tilde{\phi}^B$ are normalized versions of ${\phi}^E$ and ${\phi}^B$ and, when expressed in their Fourier-Bessel representations (see \secref{sec:fourier-bessel}), are related by
\eqn{
	\tilde{\phi}^E_{\ell m}(k) = (N_{\ell,2})^2 {\phi}^E_{\ell m}(k), \\
	\tilde{\phi}^B_{\ell m}(k) = (N_{\ell,2})^2 {\phi}^B_{\ell m}(k), 
}
where $N_{\ell, 2}=\sqrt{\frac{(\ell+2)!}{(\ell-2)!}} = (N_{\ell, -2})^{-1}$. The shear induced by gravitational lensing produces an E-mode signal only since density (scalar) perturbations cannot induce a parity odd B-mode component \cite{Castro:2005bg}.  In the absence of systematic effects (see, \eg Refs.~\cite{Bartelmann:1999yn, Kilbinger:2014cea}), $\phi^E = \phi$ and $\phi^B=0$.  The 3D cosmic shear formalism nevertheless includes B-modes since the B-mode signal---and EB-mode cross correlations---computed from data can be used as a null-test to search for residual systematics. 

The spin $\pm2$ cosmic shear signal ${}_{\pm 2}\gamma$ can also be represented in Fourier-Bessel space, which leads to a simple harmonic connection between the shear and the lensing potential.  Since shear is a spin quantity, it is decomposed into its spin Fourier-Bessel coefficients (see \secref{sec:fourier-bessel}), denoted ${}_{\pm 2}\gamma_{\ell m}(k)$.  Scalar harmonic E- and B-mode signals may be computed from ${}_{\pm 2}\gamma_{\ell m}(k)$ by
\eqn{
	E_{\ell m}(k) = -\bigl( {}_{2}\gamma_{\ell m}(k) + {}_{- 2}\gamma_{\ell m}(k)\bigr) /2, \label{eq:emode}\\
	B_{\ell m}(k) = \im\bigl( {}_{2}\gamma_{\ell m}(k) - {}_{- 2}\gamma_{\ell m}(k) \bigr) /2, \label{eq:bmode}
}
where yet another normalization is assumed (as is standard practice).  It also follows that
\eqn{
	{}_{\pm2}\gamma_{\ell m}(k) = - \bigl(E_{\ell m}(k) \pm {\rm i} B_{\ell m}(k)\bigr).
}
By \equrefs{eq:emode}{eq:bmode} and either \equrefs{eq:shearp2}{eq:shearm2} or \equrefs{eq:emodephi}{eq:bmodephi}, it follows that the lensing potential and the shear are related in harmonic space indirectly by
\eqn{
	\phi^E_{\ell m}(k) = - 2 N_{\ell, -2} E_{\ell m}(k),\\
	\phi^B_{\ell m}(k) = - 2 N_{\ell, -2} B_{\ell m}(k).
}
Consequently, the shear can be related directly to the lensing potential in Fourier-Bessel space by
\eqn{
	{}_{2}\gamma_{\ell m}(k) = \frac{1}{2} N_{\ell, 2}
	\phi^E_{\ell m}(k). \label{eq:sheartophi}
}
In fact, \equref{eq:sheartophi} can be recovered directly from the harmonic representation of \equref{chidef}.  Nevertheless, we expose the E- and B-mode decomposition since it is the standard approach and is of additional practical use in studying residual systematics, as outlined.

Finally, note that galaxy distances are not directly measurable; only their redshifts can be estimated from galaxy colours or spectra. Thus, the 3D shear signal ${}_{\pm2}\gamma(r)$ in real space is typically constructed from redshift space observations via the relation $r(z)$, which assumes a fiducial reference cosmological model \cite{Kitching:2014dtq}.  

% The spin $\pm2$ cosmic shear signal ${}_{\pm 2}\gamma$ is defined on the space formed by the product of angular position on the sky $\ang$ and comoving distance $r$; hence, its angular component can be expanded using spin spherical harmonics ${}_{\pm 2}Y_{\ell m}$ by
% \eqn{
% 	{}_{\pm 2} \gamma(\ang, r) = \gamma_1(\ang,r) \pm\im \ \gamma_2(\ang,r) = \sum_{\ell m} {}_{\pm 2}\gamma_{\ell m}(r) \ {}_{\pm 2}Y_{\ell m}(\ang).
% }
% E and B-modes signals can be computed by
% \eqn{
% 	E_{\ell m}(r) = -\bigl( {}_{2}\gamma_{\ell m}(r) + {}_{- 2}\gamma_{\ell m}(r)\bigr) /2,\\
% 	B_{\ell m}(r) = \im\bigl( {}_{2}\gamma_{\ell m}(r) - {}_{- 2}\gamma_{\ell m}(r) \bigr) /2.
% }

% from which it also follows that 

% By noting the relation ${}_2 \gamma = \frac{1}{2} (\eth\eth)\phi$ of \equref{chidef}, and as a result of the action of the $\eth$ and $\bar{\eth}$ operators on the spherical harmonics (see \appref{app:spinsha}), the lensing potential can be related indirectly to the shear by  
% \eqn{
% 	\phi^E_{\ell m}(r) = - 2 N_{\ell, -2} E_{\ell m}(r),\\
% 	\phi^B_{\ell m}(r) = - 2 N_{\ell, -2} B_{\ell m}(r),
% }
% where $N_{\ell, 2}=\sqrt{\frac{(\ell+2)!}{(\ell-2)!}} = (N_{\ell, -2})^{-1}$. Finally, these expressions may be converted into Fourier-Bessel components $\phi^E_{\ell m}(k)$ and $\phi^B_{\ell m}(k)$ using the spherical Bessel transform (defined in \appref{app:spinsha}). Similarly, shear the ${}_{\pm 2}\gamma$ may also be converted into Fourier-Bessel coefficients ${}_{\pm2}\gamma_{\ell m}(k)$. 

%==============================================================================
\subsection{Cosmology}
\label{sec:weak_lensing:cosmology}

\equrefs{eq:newton}{eq:phi} show that the lensing potential contains information about the matter fluctuations and geometry of the Universe, via $\delta$ and $F_K$, respectively. In the former, most of the information arises from the first nonzero moment of the field, the 2-point covariance  of the the fluctuations, described by the matter power spectrum $P(k; r)$:
\equ{
	\langle \ \delta_{\ell m}(k;r) \ \delta^*_{\ell' m'}(k';r) \ \rangle = P(k; r) \delta^{\rm K}_{\ell \prim{\ell}}\delta^{\rm K}_{m \prim{m}}
\delta^{\rm D}(k-k'),
}
where the angle-brackets (without a bar) indicate an ensemble average over several realizations of the field given the same cosmological model and power spectrum $P(k; r)$, $\delta^{\rm D}$ denotes the 1D Dirac delta function and $\delta^{\rm K}$ the Kronecker delta symbol. Note that we have assumed homogeneity and isotropy and that the field is evaluated at a fixed comoving distance \cite{Castro:2005bg}. Higher order statistics of the density field are of less interest 
because they are more difficult to model and measure in data. For this reason we focus on the 2-point power spectrum of the cosmic shear and other weak lensing observables.

The 3D lensing power spectrum $C^{\phi\phi}_\ell(k,\prim{k})$ is defined by
\equ{
	\langle \ \phi^E_{\ell m}(k) \ \phi^{E*}_{\ell' m'}(k')\ \rangle= C^{\phi\phi}_\ell(k,\prim{k}) \delta^{\rm K}_{\ell \prim{\ell}}\delta^{\rm K}_{m \prim{m}},
}
and is dependent on $P(k; r)$ and $F_K$.  Note that the 3D lensing potential is not homogenous and isotropic since it is given by a 2D projection of the Newtonian potential between an observer and source at radius $r$.  Consequently, the lensing potential is homogenous and isotropic in the angular direction but not in the radial direction \cite{Castro:2005bg}.  

By \equref{eq:sheartophi}, and equivalently by \equref{chidef} for $s=2$, the covariance of the 3D cosmic shear ${}_2\gamma$ in Fourier-Bessel space is related to the lensing power spectrum by
\begin{equation}
\langle \ {}_2\gamma_{\ell m}(k)\ {}_2\gamma^*_{\prim{\ell}\prim{m}}(\prim{k}) \ \rangle \ = \ \frac{1}{4} (N_{\ell, 2})^2 C^{\phi\phi}_\ell(k,\prim{k}) \delta^{\rm K}_{\ell \prim{\ell}}\delta^{\rm K}_{m \prim{m}}. \label{shearpowerspectrum}
\end{equation}
Note that since the shear transform coefficients are complex quantities, their joint probability distribution must include the correlation between the real and imaginary parts \cite{Picinbono:1996aa, Neeser:1993aa}. This may be accounted for by creating an `Affix' covariance, as shown in Ref.~\cite{Kitching:2014dtq}, or by adopting real spherical harmonics.

Further observational aspects such as the inclusion of photometric redshift information or a varying number-density of sources can also be included as shown in Ref.~\cite{Kitching:2010wa}. A further complicating aspect is that the observed galaxy ellipticity is not a direct measure of the shear but is a combination of the shear and the intrinsic (unlensed) ellipticity of the galaxy: $e^{\rm obs}(\ang,r)=e^{\rm I}(\ang,r)+\gamma(\ang,r)$, to linear order. These effects can be taken into account by modeling  auto-correlations between the intrinsic ellipticities $e^{\rm I}(\ang,r)$ and cross-correlations between the intrinsic ellipticities and the shear as described in Ref.~\cite{Kitching:2014lga} (correlations with CMB lensing can also be included, as also shown in Ref.~\cite{Kitching:2014lga}). In this paper we are concerned with linking observable power spectra to the model power spectrum that should be assumed to include all of these effects. However, for simplicity and clarity we will refer to only the `shear' power spectrum and keep the notation $C^{\phi\phi}_\ell(k,\prim{k})$. For more details on the additional modeling aspects we refer the reader to the references provided. 

We do not consider higher-order statistics, such as the bi-spectrum or the tri-spectrum \cite{Kitching:2014lga}. These can be expressed as generalizations of the power spectrum method presented. Note that in the Fourier-Bessel transforms we have assumed a flat geometry, and that the more general case requires the use of ultra spherical Bessel functions, where the geometry is now a generalized manifold. However, as pointed out in Ref.~\cite{Kitching:2006mq}, the use of the spherical Bessel functions is well motivated in the limit of $\ell \gg 1$ and $k\gg$ (curvature scale)$^{-1}$.

The standard approach of 3D cosmic shear, as performed in Ref.~\cite{Kitching:2014dtq}, is to measure the Fourier-Bessel coefficients of the shear ${}_2\gamma_{\ell m}(k)$ from a galaxy survey and then use them to constrain $C^{\phi\phi}_\ell(k,\prim{k})$ (hence, $P(k; r)$ and $F_K$) through \equref{shearpowerspectrum}. This harmonic-space approach can be challenging due to both observational and technical complications. As mentioned previously, it is difficult to simultaneously deal with complicated masks, mode coupling and scale mixing of the Fourier-Bessel coefficients. We will come back to these issues in \secref{sec:weaklensingapplication} when we detail our new method to constrain $C^{\phi\phi}_\ell$ from flaglet coefficients.

We  now step back from weak lensing and adopt a more general signal processing viewpoint to define the novel 3D spin Fourier-Laguerre and flaglet transforms (which are suitable for the analysis of arbitrary 3D spin fields). In \secref{sec:weaklensingapplication} we apply these transforms specifically to weak lensing observables and construct an estimator related to the 3D lensing power spectrum.

%==============================================================================
\section{Spin Fourier-Laguerre transform}\label{sec:flag}

After concisely reviewing signals and transforms defined on the radial line $\Rplus$, sphere $\sphere$, and rotation group SO(3), we construct an exact, separable Fourier-Laguerre transform to analyze signals of arbitrary spin defined on $\sphere\negsp{1}\times\Rplus$, \ie the 3D space formed by the product of the sphere $\sphere$ and the radial line $\Rplus$.  To subsequently construct a directional wavelet transform in this space (presented in \secref{sec:flaglet}), it is necessary to also consider signals and transforms defined on SO(3)$\times\Rplus$, \ie the space formed by the production of the rotation group SO(3) and the radial line $\Rplus$.  We thus also construct a Wigner-Laguerre harmonic transform on SO(3)$\times\Rplus$. Sampling theorems and fast algorithms to compute these transform exactly for bandlimited signals are presented.  In addition, the translation and convolution operators needed to construct a 3D spin wavelet transform are outlined. Finally, we derive the relation between the Fourier-Laguerre and Fourier-Bessel transforms, allowing one to exactly compute the Fourier-Bessel coefficients of signals bandlimited in Fourier-Laguerre space. This property is of use in \secref{sec:weaklensingapplication} to connect the Fourier-Laguerre and  flaglet transforms with the 3D weak lensing formalism. 
Note that in what follows brackets with a separating bar denote inner products (and not ensemble averages). In other words, $\langle f | g \rangle = \int_S \d\upsilon f g^*$, where the integral runs over the space $S$ of interest, \ie where $f$ and $g$ are defined, with measure $\d\upsilon$.  

%==============================================================================
\subsection{Transforms on $\Rplus$, $\sphere$, and SO(3)}

We first consider a square integrable complex signal on the radial line $f\in L^2(\Rplus)$, which can be expanded using the spherical Laguerre transform introduced in Ref.~\cite{Leistedt:2012zx}, with forward and inverse, 
\begin{eqnarray}
	{f}_p &=& \langle f | K_p \rangle = \int_{\mathbb{R}^+} \negsp{2} {\rm d} r r^2 f(r) K^*_p(r), \label{laguanalysis}\\
	f(r) &=& \sum_{p=0}^\infty {f}_p \ K_p(r) \label{lagusynthesis},
\end{eqnarray}
respectively.  The Laguerre basis functions are specified by 
\begin{equation}
	K_p(r) = \sqrt{ \frac{p!}{(p+\alpha)!} }  \frac{ e^{-{r}/{2\tau}} }{ \sqrt{\tau^{\alpha+1}}} r^{\frac{\alpha-2}{2}} L^{(\alpha)}_p\left(\frac{r}{\tau}\right),
  \label{eq:laguerrebasis}
\end{equation}
where $L^{(\alpha)}_p$ is the $p$-th generalized Laguerre polynomial of order $\alpha$ {(see \appref{app:spinsha} for further details about the basis functions $K_p$)}. In this paper we consider the case $\alpha=0$. Note that $\tau \in \mathbb{R}^+$ is a scaling factor to map the $\Rplus$ sampling theorem to any finite interval $[0,R]$ of interest. We refer the reader to \cite{Leistedt:2012zx} for more details on the spherical Laguerre transform. We take $r^2\d r$ as the natural measure on $\Rplus$ since we aim to construct 3D transforms in spherical coordinates, where the volume-invariant measure involves $ r^2\d r$ (in any case, an alternative measure could be adopted if it were desired).

We now consider a square integrable complex function on the sphere ${}_sf(\ang) \in L^2(\sphere)$ with $\ang=(\theta,\phi)$, where $\theta\in[0,\pi]$ denotes colatitude and $\phi\in[0,2\pi)$ longitude.  A spin function ${}_sf$ with spin number $s\in\mathbb{Z}$ transforms under a local rotation by ${\vartheta} \in [0,2\pi)$ in the tangent plane centered on $\ang$ as ${}_sf^\prime(\ang) = {\rm e}^{-is{\vartheta}}{}_sf(\ang)$, where the prime denotes
the rotated function \cite{Goldberg:1966uu, Newman:1966ub, Zaldarriaga:1996xe, Kamionkowski:1996ks}. For $s=\pm 2$, such as the cosmic shear field, ${}_sf$ is invariant under local rotations of $\pm \pi$.  Note that the sign convention adopted for the argument of the complex exponential differs from the original definition \cite{Newman:1966ub} but is identical to the convention used in the context of the polarization of the CMB \cite{Zaldarriaga:1996xe, Kamionkowski:1996ks}. 
The natural harmonic transform on the sphere that accounts for spin symmetry is the spin spherical harmonics transform, with forward and inverse,
 \eqn{
	{}_sf_{\ell m} &=&  \langle {}_sf | {}_{s}Y_{\ell m} \rangle = \int_{\sphere} \d\Omega(\ang) \ {}_sf(\ang) \ {}_{s}Y^*_{\ell m}(\ang), \label{spinshaanalysis}\\
	{}_sf(\ang) &=& \sum_{\ell =0}^\infty \sum_{m=-\ell}^\ell {}_sf_{\ell m}\ {}_{s}Y_{\ell m}(\ang) \label{spinshasynthesis},
}
respectively, where ${}_{s}Y_{\ell m}$ denotes the spin spherical harmonics (using
the Condon-Shortley phase convention; see \appref{app:spinsha} for more details). The usual invariant measure of the sphere is $\d\Omega(\ang)=\sin\theta \d\theta \d\phi$. Note that spin signals have ${}_sf_{\ell m}=0, \ \forall \ell < |s|$. 

The spherical Laguerre and spherical harmonics transforms can be combined naturally into a 3D separable transform, as shown in Ref.~\cite{Leistedt:2012zx} for the spin-$0$ case. Before we extend this construction to higher spin numbers in the next section, we turn to the rotation group, SO(3), and its standard harmonic transform: the Wigner transform \cite{Varshalovich:1988ye,McEwen:2015so3}. The rotation group is the natural manifold on which to construct a spherical wavelet transform probing directional features. This is due to the rotation properties of spherical harmonics and the natural convolution operator on the sphere (recalled in \secref{sec:convolutions}). Thus, we consider a square integrable complex signal defined on the rotation group $f(\rhoang)\in L^2$(SO(3)), with $\rhoang=(\alpha,\beta,\gamma)$, where $\alpha\in[0,2\pi), \beta\in[0,\pi], \gamma\in[0,2\pi)$ are the Euler angles. Its forward and inverse Wigner transforms are given by 
\eqn{
	f^\ell_{ m n} &=&  \langle f | D^{\ell *}_{m n} \rangle = \int_{\rm SO(3)} \negsp{3} \d\mu(\rhoang) \ f(\rhoang) \ D^{\ell}_{m n}(\rhoang), \label{wigneranalysis}\\
	f(\rhoang) &=& \sum_{\ell =0}^\infty \frac{2\ell+1}{8\pi^2} \sum_{m=-\ell}^\ell \sum_{n=-\ell}^{\ell} f^{\ell}_{m n}\ D^{\ell*}_{m n}(\rhoang) \label{wignersynthesis},
}
respectively, where $D^{\ell}_{m n}$ are Wigner functions and the invariant measure on the rotation group reads $\d\mu(\rhoang)=\sin\beta \ \d\beta \d\alpha \d\gamma$ (see \appref{app:spinsha} for further details). 

% Note that spin signals can be described on the rotation group as the Wigner functions are related to the spin spherical harmonics through \cite{Goldberg:1966uu}
% \eqn{
% 	{}_sY_{\ell m}(\theta,\phi) = (-1)^s \sqrt{\frac{2\ell+1}{4\pi}} D^{\ell*}_{m,-s}(\phi, \theta, 0) \label{sshttowigner}.
% } 

%==============================================================================
\subsection{Fourier-Laguerre transform on $\sphere\negsp{1}\times\Rplus$ and Wigner-Laguerre transform on SO(3)$\times\Rplus$}

\begin{figure*}
\centering
\setlength{\unitlength}{1cm}
\hspace*{-5mm}\includegraphics[height=5.3cm, trim = 5cm 8.cm 5.cm 8.cm, clip]{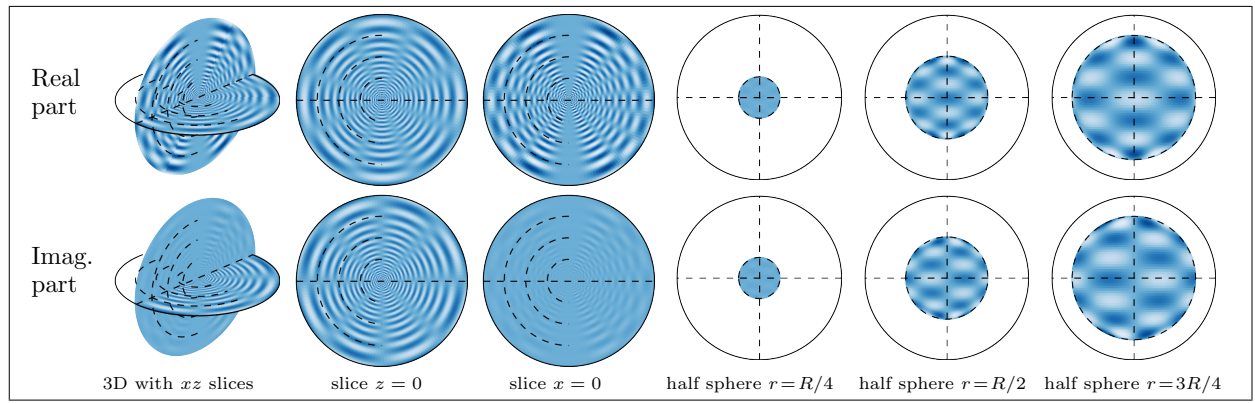}
\caption{Real and imaginary parts of the Fourier-Laguerre basis function ${}_sY_{\ell m}(\ang)K_p(r)$ with $s=0,\ \ell=10,\ m=5,\ p=40$, shown on the 3D Fourier-Laguerre sampling scheme with $L=P=80$, plotted up to $r=R/2$ with $R$ the radius of the final sample of the radial sampling scheme \cite{Leistedt:2012zx}. This plot shows that the 3D basis functions successfully probe harmonic modes in the angular and radial directions separately.  Furthermore, notice that the basis functions are not well localized in space, similar to standard Fourier bases.}\label{fig:flagbasis}
\end{figure*}

We now combine the previous transforms and extend the Fourier-Laguerre tranform presented in Ref.~\cite{Leistedt:2012zx} so that its angular part can support spin signals on $\sphere$, in order to match the nature of weak lensing observations. More precisely, we now consider 3D complex functions in spherical coordinates ${}_sf(\ang,r)\in L^2(\sphere\negsp{1}\times\Rplus)$.  We create a separable transform by combining the spherical Laguerre transform with the spin spherical harmonics, leading to the forward and inverse transforms
\begin{eqnarray}
	{}_sf_{\ell m, p} &=& \langle \ {}_sf \ | \ {}_sY_{\ell m} K_p \ \rangle
	 =  \int_{\sphere}  \d\Omega(\ang) \int_{\Rplus} \negsp{2}  \d r r^2  {}_sf(\ang,r)  \ {}_sY^*_{\ell m}(\ang)  K^*_p(r), \label{flags2analysis} \\
	{}_sf(\ang,r) &=& \sum_{\ell m p} \ {}_sf_{\ell m, p} \ {}_sY_{\ell m}(\ang) \ K_p(r)\label{flags2synthesis},
\end{eqnarray}
respectively. The spin property implies that ${}_sf_{\ell m,p}=0, \ \forall \ell < |s|$. 

For functions $f(\rhoang,r)\in L^2($SO(3)$\times\Rplus)$ we define the Wigner-Laguerre forward and inverse transforms as
\begin{eqnarray}	
	f^\ell_{m n, p} &=&  \langle \ f \ | \ D^{\ell*}_{m n} K_p \ \rangle 
	= \int_{\rm SO(3)} \negsp{5} \d\mu(\rhoang) \int_{\Rplus} \negsp{2}  \d r r^2  f(\rhoang,r)   D^\ell_{m n}(\rhoang)  K^*_p(r) \label{flagso3analysis} ,\\
	f(\rhoang,r) &=& \sum_{\ell m n p} \frac{2\ell+1}{\ 8\pi^2} \ f^\ell_{m n, p} \ D^{\ell*}_{m n}(\rhoang) \ K_p(r),\label{flagso3synthesis}
\end{eqnarray}
respectively. In these expressions, the $\ell m n$ indices refer to the angular modes (in the spherical harmonics or Wigner transforms), 
while $p$ corresponds to the radial mode (from the spherical Laguerre transform). An example of a basis function is shown 
in \figref{fig:flagbasis}. Finally, as in the individual transforms, the summations for the harmonic modes 
run over $p\in\mathbb{N}$, $\ell\in\mathbb{N}$, $-\ell\leq m \leq\ell$ and $-\ell\leq n \leq\ell$. In what follows, 
we will omit these bounds for conciseness and use the terminology `angular' for both the $\sphere$ and SO(3) 
parts of the signals and transforms.

%==============================================================================
\subsection{Sampling theorems and numerical implementation}

In practice, evaluating the integrals in \equrefs{flags2analysis}{flagso3analysis} numerically requires quadrature rules. 
While generic numerical integration methods can be adopted, this problem can be tackled more accurately by appealing to sampling theorems. When considering bandlimited signals, 
sampling theorems lead to exact quadrature rules, which allow one to discretize the space such that the previous 
integrals can be computed exactly. In practice, signals are transformed with accuracy close to the level of machine precision, with errors due to the (finite, approximate) representation of floating point numbers only. This approach is particularly desirable when performing numerous transforms successively or when high numerical accuracy is required (which is typically the case in modern observational cosmology).

The radial and angular parts of the Fourier-Laguerre and Wigner-Laguerre transforms are fully separable. This implies that one can simply 
combine sampling theorems on the sphere, the rotation group and the radial line to obtain sampling 
theorems on $\sphere\negsp{1}\times\Rplus$ and SO(3)$\times\Rplus$. We now recall these sampling theorems and 
the notation for the sampling points and quadrature weights.

On $\Rplus$, we use the spherical Laguerre quadrature rule presented in Ref.~\cite{Leistedt:2012zx}. bandlimited signals have $f_p = 0,\ \forall p>P$, with $P$ the radial band-limit, while the sampling nodes and quadrature weights are denoted by $r_k$ and $w_k$, respectively. This exact quadrature, with corresponding nodes and weights, is used to compute \equref{laguanalysis}. 

On the sphere, we consider the sampling theorem developed in Ref.~\cite{McEwen:2011em}: 
a function ${}_sf$ with 
band limit $L$ satisfies ${}_sf_{\ell m}=0,\ \forall \ell\geq L$ and can be completely 
described by the values taken on the equiangular, separable pixelization: $\ang_{ij} = (\theta_i, \phi_j)$ 
with $i=0,\dots,L-1$ and $j=0,\dots,2L-1$.  The spherical harmonics coefficients of \equref{spinshaanalysis} can 
be calculated with a discrete sum of ${}_sf(\ang_{ij})$ with quadrature weights $w_{ij}$. 
In principle, other sampling theorems could be used (\eg Refs.~\cite{Crittenden:1998jm, Doroshkevich:2003xb}), but note that at fixed band limit this sampling scheme requires the fewest samples \cite{McEwen:2011em}.  pixelization schemes not based on sampling theorems could also be adopted (\eg HEALPix \cite{Gorski:1999rt}), in which case larger errors due to approximate numerical quadrature are expected. {On SO(3), a function $f$ with angular band limit $L$ and azimuthal band limit $N$ satisfies $f^\ell_{ m n}=0,\ \forall \ell\geq L,\  |n|>N$. }
%n=-N,\dots,N,\ m=-\ell,\dots,\ell$. 
The sampling theorem of Ref.~\cite{McEwen:2011em} was extended to SO(3) in Ref.~\cite{McEwen:2015so3}: the sampling points, also equiangular and separable, are denoted by $\rhoang_{hij} = (\alpha_i, \beta_j, \gamma_h)$ with $i=0,\dots,L-1$, $j=0,\dots,2L-1$, and $h=0,\dots,2N-1$. The quadrature weights are denoted by $w_{hij}$ and can be used to compute the Wigner coefficients of \equref{wigneranalysis} exactly.

The theoretically exact spin Fourier-Laguerre transform for bandlimited signals on $\sphere\negsp{1}\times\Rplus$ reads
\eqn{	
	{}_sf_{\ell m, p} &=& \sum_{ijk} \ w_{ij} \   w_k \  {}_sf(\ang_{ij},r_k)  \ {}_sY^*_{\ell m}(\ang_{ij})  \ K_p(r_k),  \ \ \label{eq:flag_s2_discrete} \\
	{}_sf(\ang_{ij},r_k) &=& \sum_{\ell m p} \ {}_sf_{\ell m, p} \ {}_sY_{\ell m}(\ang_{ij}) \ K_p(r_k),
}
which are the discrete versions of \equrefs{flags2analysis}{flags2synthesis}, respectively.

The theoretically exact Wigner-Laguerre transform for bandlimited signals on SO(3)$\times\Rplus$ reads
\eqn{	
	f^\ell_{m n, p} &=& \sum_{hijk} \ w_{hij} \  w_k\ f(\rhoang_{hij},r_k) \ D^\ell_{m n}(\rhoang_{hij})   \    K_p(r_k), \ \ \label{eq:flag_so3_discrete} \\
	f(\rhoang_{hij},r_k) &=& \sum_{\ell m n p} \ \frac{2\ell+1}{\ 8\pi^2} \  f^\ell_{m n, p} \ D^{\ell*}_{m n}(\rhoang_{hij}) \ K_p(r_k),
}
which are the discrete versions of \equrefs{flagso3analysis}{flagso3synthesis}, respectively.

{Although we express the forward transforms of \equrefs{eq:flag_s2_discrete}{eq:flag_so3_discrete} using discrete quadrature above, in practice we do not compute these expressions explicitly but rather compute them using fast algorithms, drawing on Refs.~\cite{McEwen:2011em, McEwen:2015so3}, as discussed below. The sums in these equations are finite and adjusted to the band limits (for the $\ell m n p$ indices) and to the sampling nodes (for the $hijk$ indices). }
%Letting $L$, $N$ and $P$ denote the angular, azimuthal, and radial band limits, respectively, the bounds for these indices are $\ell=0,\dots,L-1$, $m=-\ell,\dots,\ell$, $n=-N,\dots,N$, $p=0,\dots,P-1$, $i=0,\dots,L-1$, $j=0,\dots,2L-1$, and $h=0,\dots,2N-1$, $k=0,\dots,P-1$. 
In what follows we will omit the pixel indices and bounds of the summations for conciseness. We will also use the integral (continuous) forms of the transforms for clarity, since the sampling theorem guarantees that these can be evaluated exactly for bandlimited signals (which is typically the case in practical applications).

As described above, any bandlimited signal with radial and angular band limits $P$ and $L$, respectively, can be represented exactly by $\sim 2PL^2$ samples on $\sphere\negsp{1}\times\Rplus$ thanks to the combination of the sampling theorem for the pixelization of Ref.~\cite{McEwen:2011em} and the spherical Laguerre sampling theorem \cite{Leistedt:2012zx}. Similarly, signals on the rotation group, with the same band limits $P$ and $L$, and an additional azimuthal 
band limit $N$, are captured by $\sim 4PNL^2$ samples on SO(3)$\times\Rplus$, thanks also to the corresponding sampling theorem on the rotation group \cite{McEwen:2015so3}.

In terms of computational complexity, the separability of the angular and radial components in the Fourier-Laguerre and Wigner-Laguerre transforms is an essential property. Fast algorithms exist to compute transforms on the sphere and the rotation group \cite{Driscoll:1994cca, mcewen:2006:fcswt, Wiaux:2005fm, McEwen:2011em, McEwen:2013tpa, McEwen:2015so3}. In particular, we use the implementations detailed in Refs.~\cite{McEwen:2011em,McEwen:2015so3}, which sets the complexity of the spin spherical harmonics and Wigner transforms  to $\mathcal{O}(L^3)$ and $\mathcal{O}(NL^3)$, respectively, by exploiting fast Fourier transforms on the ring torus.  The spherical Laguerre transform simply 
scales as $\mathcal{O}(P)$, and the quadrature weights and basis functions are computed through 
recurrence relations \cite{Leistedt:2012zx}. Therefore, the spin Fourier-Laguerre and Wigner-Laguerre transforms scale as $\mathcal{O}(PL^3)$ and $\mathcal{O}(NPL^3)$, respectively. Note that the spin number $s$ is a parameter that does not change the complexity or the accuracy of any of the transforms, which is often not the case for alternative approaches.

%==============================================================================
\subsection{Connection to the Fourier-Bessel transform}
\label{sec:fourier-bessel}

In contrast to the Fourier-Laguerre construction, the standard nonseparable basis for 
3D spin functions ${}_sf(\ang,r)\in L^2(\sphere\negsp{1}\times\Rplus)$ (such as weak lensing observables) is the 
combination of spherical harmonics with spherical Bessel functions $j_\ell$. These define the Fourier-Bessel transform, 
already used in \secref{sec:weaklensingintro}, with forward and inverse 
\begin{eqnarray}
	{}_sf_{\ell m}(k) &=& \langle \ {}_sf \ | \ {}_sY_{\ell m} j_\ell(k\cdot) \ \rangle 
	 = \int_{\sphere}  \d\Omega(\ang) \sqrt{\frac{2}{\pi}}  \int_{\Rplus} \negsp{2}  \d r r^2  {}_sf(\ang,r)  \ {}_sY^*_{\ell m}(\ang)  j^*_\ell(kr),\label{fourierbesselanalysis} \\
	{}_sf(\ang,r) &=&  \sum_{\ell m} \sqrt{\frac{2}{\pi}} \int_\Rplus\negsp{2} \d k k^2 \ {}_sf_{\ell m}(k)  \ {}_sY_{\ell m}(\ang)   j_\ell(kr) \label{fourierbesselsynthesis} .
\end{eqnarray}

Unlike in the Fourier-Laguerre case, the Fourier-Bessel basis functions and transform are not separable: the 
angular modes $\ell$ are coupled with the radial modes $k$ through the spherical Bessel 
function $j_\ell(kr)$. This is because the basis functions are solutions of the (isotropic) Laplacian 
operator in spherical coordinates, with eigenvalues $-k^2$. It is important to note that the Fourier-Bessel 
transform is notoriously difficult to evaluate for generic functions because it does not admit a 
sampling theorem \cite{McEwen:2013jpa, Leistedt:2012zx}. Therefore, one must appeal to
an approximate quadrature rule for the radial integrals (see \eg Refs.~\cite{Heavens:2003jx, Leistedt:2011mk, Lanusse:2011cm}), which is challenging because the spherical Bessel 
functions have infinite support and oscillate rapidly.

Exact analytical formula only exist for simple forms of signals, but the first exact formula to compute 
the Fourier-Bessel transform of a wide class of signals was presented in Ref.~\cite{Leistedt:2012zx}. The starting 
point of this approach is to decompose the spherical Bessel functions in the spherical Laguerre basis, \ie
\begin{eqnarray}
	j_\ell(kr) &=& \sum_{p=0}^\infty \mathcal{J}_{\ell, p}(k) K_p(r), \\
	\mathcal{J}_{\ell, p}(k) &=& \int_{\Rplus}\negsp{2} \d rr^2 j_\ell(kr) K^*_p(r).\label{besselflagcoefs} 
\end{eqnarray}
The Fourier-Bessel coefficients of ${}_sf$ can then be expressed in terms of its Fourier-Laguerre coefficients:
\begin{eqnarray}
	{}_sf_{\ell m}(k) = \sqrt{\frac{2}{\pi}}  \sum_p \ {}_sf_{\ell m, p}  \ \mathcal{J}_{\ell, p}(k) \label{besselflagconnection}.
\end{eqnarray}
This sum is finite if ${}_sf$ is bandlimited in spherical Laguerre space, and ${}_sf_{\ell m, p}$ is also evaluated exactly thanks to the exact quadrature rules of the Fourier-Laguerre transform. Thus, \equref{besselflagconnection} provides a way to compute the Fourier-Bessel transform of signals described in Fourier-Laguerre space exactly. It was also shown in Ref.~\cite{Leistedt:2012zx} that $\mathcal{J}_{\ell, p}(k)$ admits an analytical form, involving the moments of $j_\ell(kr)$. Although $\mathcal{J}_{\ell, p}(k)$ can still be challenging to evaluate numerically for high $\ell$, it does not depend on the signal ${}_sf$ under consideration and can be tabulated. We will see that $\mathcal{J}_{\ell, p}(k)$ appears in the treatment of weak lensing observables, natively expressed in the Fourier-Bessel basis.

%==============================================================================
\subsection{Rotation, translation, and convolutions}\label{sec:convolutions}

We now construct rotation and convolution operators on $\sphere\negsp{1}\times\Rplus$  and SO(3)$\times\Rplus$, generalizing the operators of the scalar Fourier-Laguerre transform \cite{McEwen:2013jpa, Leistedt:2012zx}. {In general, such operators are essential for constructing a meaningful wavelet transform, where the wavelet coefficients of a signal are its convolution with the wavelets. Since wavelets have compact support in real and frequency space, this approach allows the transform to extract well defined scales and directions.}

We first recall that the rotation of a spin function on the sphere ${}_sf\in L^2(\sphere)$ is naturally defined by 
\eqn{
	(\mathcal{R}_\rhoang \ {}_sf)(\ang)= {}_sf(\mathrm{R}^{-1}_\rhoang{\bf x}),
}
where ${\bf x}$ is the Cartesian vector corresponding to $\ang$ and $\mathrm{R}^{-1}_\rhoang$ is the 3D rotation matrix corresponding to the rotation operator $\mathcal{R}_\rhoang$.  In harmonic space, the rotation conveniently reduces to \cite{McEwen:2015s2let}
\eqn{
	(\mathcal{R}_\rhoang \ {}_sf)_{\ell m} = \sum_{n=-\ell}^{\ell} D^\ell_{mn}(\rhoang) \ {}_sf_{\ell n}.
}
Furthermore, we define the translation of a radial function $f\in\Rplus$ in spherical Laguerre space as \cite{Leistedt:2012zx, McEwen:2013jpa}
\eqn{
	(\mathcal{T}_r f )_p = f_p K_p(r).
}
This is analogous to the harmonic representation of the translation operator for functions on the infinite (Cartesian) line $\mathbb{R}$ (but not analogous to the real space expression of the translation operator).  In addition, such a translation operator can be expressed in terms of convolution with a shifted Dirac delta function, again in analog to translation on $\mathbb{R}$, as detailed in Ref.~\cite{McEwen:2013jpa}.

We now consider functions on $\sphere\negsp{1}\times\Rplus$, and introduce a directional convolution operator $\convdir$ such that its action on $f,g$ yields a function on SO(3)$\times\Rplus$ defined by
\eqn{
	 (f  \convdir g)(\rhoang,t)  &=& \langle  f  \ | \mathcal{R}_\rhoang \mathcal{T}_t \ g  \rangle \ = \  \int_{\sphere}  \d\Omega(\ang)  \int_{\Rplus} \negsp{2} \d r r^2\ f(\ang,r) \ ( \mathcal{R}_\rhoang \mathcal{T}_t \ g )^*(\ang, r)  .
}
We also introduce an axisymmetric convolution operator $\convaxisym$ (a special case of $\convdir$) such that its action between $f$ and an axisymmetric function $h$ (satisfying $\mathcal{R}_{(0,0,\gamma)} h = h,\ \forall \gamma$) yields a function on $\sphere\negsp{1}\times\Rplus$ defined by
\eqn{
	 (f  \convaxisym h)(\ang^\prime,t)  &=& \langle  f  \ | \mathcal{R}_{\ang^\prime} \mathcal{T}_t \ h  \rangle \ = \ \int_{\sphere}  \d\Omega(\ang)  \int_{\Rplus} \negsp{2} \d r r^2\ f(\ang,r) \ ( \mathcal{R}_{\ang^\prime} \mathcal{T}_t \ g )^* (\ang, r) ,
}
where $\mathcal{R}_{\ang^\prime}$ is the rotation operator for axisymmetric functions, \ie $\mathcal{R}_{\ang^\prime} = \mathcal{R}_{(\phi^\prime, \theta^\prime, 0)} $. In what follows,  $\mathcal{R}_\rhoang$ and $\mathcal{R}_{\ang}$  refer to directional and axisymmetric rotations, respectively, since $\rhoang$ and $\ang$ denote angles on SO(3) and $\sphere$, respectively. More detailed discussions about rotation, translation, and convolution operators on the sphere and the radial line can be found in Refs.~\cite{McEwen:2015s2let} and \cite{Leistedt:2012zx, McEwen:2013jpa}, respectively. All operators can be evaluated exactly via their harmonic representations and all transforms from pixel to harmonic space and conversely are exact thanks to the sampling theorems adopted.

%==============================================================================
%==============================================================================
\section{Spin flaglet transform}\label{sec:flaglet}

{We now extend the scalar, axisymmetric flaglet transform of Ref.~\cite{Leistedt:2012zx} to analyze signals of arbitrary spin and to probe their features in the angular direction. This new separable flaglet transform is defined in Sec.~\ref{sec:flaglet1}, where we present the core equations and rely on the previous translation and convolution operators. However, to highlight the generic properties of the transform we do not specify the details of the flaglet kernels ${}_s\Psi^{ij}$ and ${}_s\Phi^{}$ until Sec.~\ref{sec:flaglet2}, where we also present our efficient implementation using sampling theorems and multiresolution algorithms.}

%==============================================================================
\subsection{Transform}\label{sec:flaglet1}

The flaglet coefficients of a spin signal ${}_sf\in L^2(\sphere\times\Rplus)$, denoted by $W_{{}_sf}^{{}_s\Psi^{ij}}$, are defined on SO(3)$\times\Rplus$ as the directional convolution of ${}_sf$ with the flaglets ${}_s\Psi^{ij}\in L^2(\sphere\times\Rplus)$
\eqn{
	 W_{{}_sf}^{{}_s\Psi^{ij}}(\rhoang,t) &=&    ({}_sf  \convdir {}_s\Psi^{ij})(\rhoang,t) \ = \ \int_{\sphere} \d\Omega(\ang)  \int_{\Rplus} \negsp{2} \d r r^2\ {}_sf(\ang,r) \ (\mathcal{R}_\rhoang \mathcal{T}_t \ {}_s\Psi^{ij})^*(\ang, r)  \label{eq:wavanalysis1}, 
}
 where $i=I_0,\dots,I$ and $j=J_0,\dots,J$ denote the angular and radial scales, respectively. Flaglets are localized in scale, position and angular orientation, 
and designed to capture the directional, high-frequency content of the signal of interest.  {A scaling function ${}_s\Phi \in L^2(\sphere\times\Rplus)$ is introduced to capture the low frequency content of the signal in the scaling coefficients $W_{{}_sf}^{{}_s\Phi}$. In this work we analyze the low frequency part of the signal using axisymmetric flaglets because its directional content is typically of lower interest. In the context of weak-lensing, largest scales (low $k$ in $C_\ell^{\phi\phi}$) have far fewer modes, contain less information (due to the reduced amplitude of $C_\ell^{\phi\phi}$), and are more difficult to deal with due to observational masks and other considerations mentioned above. However, extending the formalism presented here to support directional scaling functions is possible and straightforward.  }

We define $W_{{}_sf}^{{}_s\Phi} \in L^2(\sphere\negsp{1}\times\Rplus)$ 
as the axisymmetric convolution of ${}_sf$ with the axisymmetric scaling function ${}_s\Phi$
\eqn{
	 W_{{}_sf}^{{}_s\Phi}(\ang^\prime,t) &=&    ({}_sf  \convaxisym {}_s\Phi)(\ang^\prime,t) \ = \ \int_{\sphere}\d\Omega(\ang)  \int_{\Rplus} \negsp{2} \d r r^2\ {}_sf(\ang,r) \ ( \mathcal{R}_{\ang^\prime} \mathcal{T}_t \ {}_s\Phi )^*(\ang, r) .
}
Since ${}_s\Phi$ is axisymmetric, the Fourier-Laguerre coefficients ${}_s\Phi_{\ell m,p}$ are nonzero for $m=0$ only. 

In what follows we will use a compressed notation for all sums over scales $i$ and $j$, assumed to run over $I_0,\ldots,I$ and $J_0,\ldots,J$, 
respectively. These bounds, as well as the flaglets and scaling functions ${}_s\Psi^{ij}$ and ${}_s\Phi^{}$, are consistently defined in the next 
section in order to achieve exact reconstruction of the input (bandlimited) signal.  

The wavelet transform makes direct use of the rotation, translation and convolution operators defined previously. Hence, 
the forward flaglet transform (or analysis step, equivalent to \equref{eq:wavanalysis1}) can also 
be expressed in Wigner-Laguerre space (on SO(3)$\times\Rplus$)  in a straightforward manner by 
\eqn{
	\left(W_{{}_sf}^{{}_s\Psi^{ij}}\right)^\ell_{m n, p} &=& \langle W_{{}_sf}^{{}_s\Psi^{ij}} | D^{\ell*}_{m n} K_p \rangle   \ = \ \frac{8\pi^2}{2\ell+1} {}_sf_{\ell m, p} \ {}_s\Psi_{\ell n, p}^{ij}{}^* \label{eq:wavanalysis_lm}.
}
The scaling coefficients (on $\sphere\times\Rplus$) in Fourier-Laguerre space read
\eqn{
	{}_s(W_{{}_sf}^{{}_s\Phi})_{\ell m, p} &=& \langle W_{{}_sf}^{{}_s\Phi} | {}_sY_{\ell m} K_p \rangle   \ = \ \sqrt{\frac{4\pi}{2\ell+1}} {}_sf_{\ell m, p} \ {}_s\Phi_{\ell m, p}{}^* \label{eq:wavanalysisscal_lm}.
}

The reconstruction of the original signal is achieved through the inverse flaglet transform (synthesis step)
\eqn{
	&&  \negsp{3} {}_sf(\ang, r) = 
	 \int_{\sphere}  \d\Omega(\ang^\prime) \negsp{1}  \int_{\Rplus} \negsp{2}  \d r r^2 \ W_{{}_sf}^{{}_s\Phi}(\ang^\prime) \ ( \mathcal{R}_{\ang^\prime} \mathcal{T}_r\ {}_s\Phi )(\ang, t)  \ + \  \sum_{ij}\int_{\rm SO(3)} \negsp{5} \d\mu(\rhoang) \negsp{1}  \int_{\Rplus} \negsp{2}  \d r r^2 \ W_{{}_sf}^{{}_s\Psi^{ij}}(\rhoang) \ ( \mathcal{R}_\rhoang \mathcal{T}_r\ {}_s\Psi^{ij} )(\ang, t)   \label{eq:wavsynthesis}, \quad\quad
}
or in Fourier-Laguerre space
\eqn{
	{}_sf_{\ell m, p} \ = \ \sqrt{\frac{4\pi}{2\ell+1}} \ {}_s\Phi_{\ell m, p} \ {}_s(W_{{}_sf}^{{}_s\Phi})_{\ell m, p}  \ + \  \sum_{ij}\sum_n  \ {}_s\Psi_{\ell n, p}^{ij} \ (W_{{}_sf}^{{}_s\Psi^{ij}})^\ell_{ m n, p} \label{eq:wavsynthesis_lm}.
	}

In order for the transform to be exact, \ie for the wavelet coefficients to capture all the information content of ${}_sf$, the flaglets and scaling function must be defined to satisfy \equrefss{eq:wavanalysis_lm}{eq:wavanalysisscal_lm}{eq:wavsynthesis_lm}. In other words, they must be chosen such that their Fourier-Laguerre coefficients satisfy the admissibility condition
\eqn{
	\frac{4\pi}{2\ell+1} |{}_s\Phi_{\ell 0, p}|^2  + \frac{8\pi^2}{2\ell+1} \sum_{ijm} | {}_s\Psi_{\ell m, p}^{ij} |^2 = 1, \ \ \ \forall \ell, p \  \label{admissibility}.
}
	
%==============================================================================
\subsection{Wavelet construction}\label{sec:flaglet2}

\begin{figure}
\setlength{\unitlength}{1cm}
\hspace*{-4mm}\includegraphics[height=5.1cm, trim = 1.5cm 11.1cm 2.cm 11.2cm, clip]{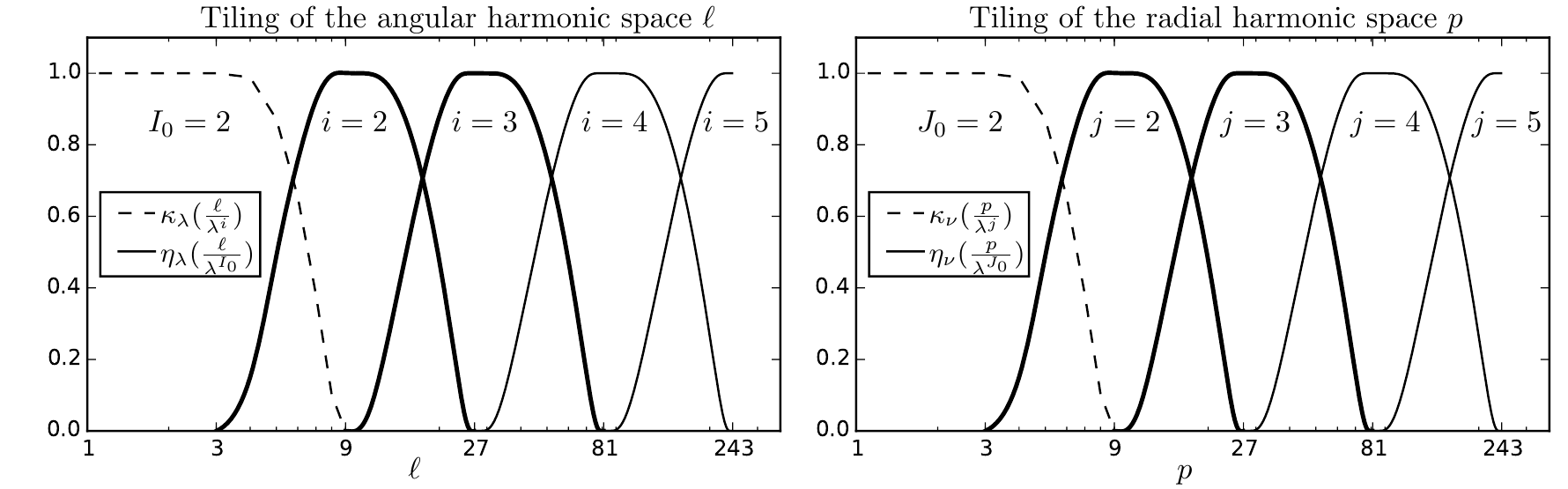}
\caption{Tiling of the angular (left panel) and radial (right panel) harmonic lines employed to construct flaglets ${}_s\Psi^{ij}$ in \equref{flaglet_lmp}. The left panel shows the kernels $\kappa_\lambda(\frac{\ell}{\lambda^i})$ and $\eta_\lambda(\frac{\ell}{\lambda^{I_0}})$ tiling the range $\ell=0,\dots,L-1$ (zero is not shown but is included in $\eta_\lambda$), while the right panel shows $\kappa_\nu$ and $\eta_\nu$ tiling $p=0,\dots,P-1$.  Since $L=P=243$ and $\lambda=\nu=3$, six kernels are needed to fully tile $[0, L-1]$ and $[0, P-1]$, \ie the scales considered are $i=0,\dots,5$ and $j=0,\dots,5$. The first two scales are incorporated in the scaling function by setting $I_0=J_0=2$. The thicker lines highlight the scales shown in \figrefs{fig:scalarflaglets}{fig:spinflaglets}. Note that we do not show the hybrid kernel $\eta_{\lambda \nu}$, included in the scaling function to satisfy the admissibility condition at low $p$ and $\ell$, see \equrefs{admissibility}{scalingfunctionfull}. We also do not include the directional component $\zeta_{\ell m}$ in this figure, which is nevertheless shown in real space in \figref{fig:scalarflaglets}.}\label{fig:tiling}
\end{figure}

So far we have constructed a generic flaglet transform relying on the properties of the Fourier-Laguerre space. To uniquely characterise this transform, we need to specify the scaling function ${}_s\Phi$, the flaglets ${}_s\Psi^{ij}$, and the bounds $I_0$, $I$, $J_0$, $J$ to satisfy \equref{admissibility}. {We follow the construction of scalar flaglets \cite{Leistedt:2012zx} and scale-discretized wavelets \cite{Wiaux:2007ri, McEwen:2015s2let}. Other types of wavelets could be used; we opted for scale-discretized wavelets because they exhibit good localization properties
\cite{mcewen:2015s2let_localisation} and can be easily extended to probe directional features \cite{McEwen:2015s2let,Wiaux:2007ri}. They exhibit a compact representation in harmonic space that also allows several computational improvements, which are essential to make a 3D spin directional transform tractable. We use tiling functions $\kappa$ and $\eta$ defined as}
\eqn{
	\kappa_\lambda(t) &=& \sqrt{ k_\lambda(t/\lambda) - k_\lambda(t) } ,\label{eq:kappadef}\\
	\eta_{\lambda}(t) &=& \sqrt{ k_\lambda(t)}, \label{eq:etadef}
}
where
\begin{equation}
  k_\lambda(t) = \frac{\int_{t}^1\frac{{\rm d}t^\prime}{t^\prime}s_\lambda^2(t^\prime)}{\int_{1/\lambda}^1\frac{{\rm d}t^\prime}{t^\prime}s_\lambda^2(t^\prime)}, \label{smoothscaling}
\end{equation}
which is unity for $t<1/\lambda$, zero for $t>1$, and is smoothly decreasing from unity to zero for $t \in [1/\lambda,1]$. In these equations, $\lambda$ is the wavelet dilation parameter and characterises the density of the tiling. In what follows, $\lambda$ will refer to the wavelet tiling in the angular direction, while a second parameter $\nu$ is introduced to tile the radial direction (with functions $\kappa_\nu$ and $\eta_\nu$ constructed in the same manner). Finally, the remaining free function $s_\lambda(t)$ (also defined for $\nu$) must have compact support in $[\frac{1}{\lambda}, 1]$  and is defined as $s( \frac{2\lambda}{\lambda-1} (t-1/\lambda)-1)$ with
\begin{equation}
	s(t) = \left\{ \begin{array}{ll} \ e^{-\frac{1}{1-t^2}}, & t\in[-1,1] \\ \  0, & t \notin [-1,1]\end{array} \right. .
\end{equation}
The tiling of the radial and angular harmonic spaces constructed with these functions is shown in \figref{fig:tiling} for $\lambda=\nu=3,\ I_0=J_0=2,\ L=P=243$.

%%%%%%%%%%%%%
%%%%%%%%%%%%%

In addition, as for the scalar axisymmetric flaglet transform \cite{Leistedt:2012zx}, a hybrid tiling function is needed in order to construct a suitable scaling function $\Phi$ and satisfy \equref{admissibility},
\begin{eqnarray}
	\eta_{\lambda \nu}(t, t^\prime) \negsp{1}=\negsp{1}  \sqrt{ k_\lambda(t/\lambda)k_\nu(t^\prime) 
		 + k_\lambda(t)k_\nu(t^\prime / \nu)  
		 -  k_\lambda(t)k_\nu(t^\prime) }. \ \ \ \label{hybridkernel}
\end{eqnarray}

With the tiling functions $\kappa$ and $\eta$ defined in \equref{eq:kappadef} and \equref{eq:etadef}, we construct the flaglets in Fourier-Laguerre space as
\eqn{
	{}_s\Psi_{\ell m, p}^{ij} = \sqrt{\frac{2\ell+1}{8\pi^2}} \ \kappa_\lambda\left(\frac{\ell}{\lambda^i}\right) \kappa_\nu\left(\frac{\phantom{\ell}\negsp{2}p}{\nu^{j}}\right)  \ \zeta_{\ell m}. \label{flaglet_lmp}
}
The extra function $\zeta$ controls the directionality component \cite{McEwen:2015so3, McEwen:2013tpa, Wiaux:2007ri} and is parameterised by an azimuthal band limit $N$, such that $\zeta_{\ell m}=0,\ \forall \ell, m$ with $|m|\geq N$. It is defined in harmonic space as 
\eqn{
	\zeta_{\ell m} = a b \sqrt{ \frac{1}{2^c} \left(\begin{array}{c} c \\ (c-m)/2 \end{array}\right) },
}
to satisfy various azimuthal symmetries and normalized such that $\sum_{m}\zeta_{\ell m}=1$, for all values of $\ell$ for which $\zeta_{\ell m}$ are nonzero for at least one value of $m$ (see, \eg, \cite{mcewen:2015s2let_localisation}). In this expression, $a=1$ for $N-1$ odd and $a=\im$ for $N-1$ even, $b=[1-(-1)^{N+m}]/2$ and $c=\min\{N-1, \ell-[1-(-1)^{N+\ell}]/2\}$.

Finally, the axisymmetric scaling function is needed to capture the information from the regions $\ell \leq \lambda^{I_0}$ and $p \leq \nu^{J_0}$, which are not probed by the flaglets defined above. Thus, it is constructed to satisfy \equref{admissibility} and reads
\begin{equation}
	{}_s{\Phi}_{\ell 0, p}  =   \left\{ \begin{array}{ll} 
	 \negsp{1} \sqrt{ \frac{2 \ell+1}{4\pi}}  \ \eta_{\nu}\left(\frac{\phantom{\ell}\negsp{2}p}{\nu^{J_0}}\right) , & \textrm{if } \ell  > \lambda^{I_0}, \ p \leq \nu^{J_0}	 \\
  \negsp{1} \sqrt{ \frac{2 \ell+1}{4\pi}}  \ \eta_{\lambda}\left(\frac{\ell}{\lambda^{I_0}}\right)  , & \textrm{if } \ell \leq \lambda^{I_0} , \ p  > \nu^{J_0}  \\
\negsp{1}  \sqrt{ \frac{2 \ell+1}{4\pi}}  \ \eta_{\lambda \nu}\left(\frac{\ell}{\lambda^{I_0}},\frac{p}{\nu^{J_0}}\right) , \quad\quad & \textrm{if }  \ell < \lambda^{I_0}, \ p  < \nu^{J_0} \\
  \quad 0 , & \textrm{elsewhere.} \end{array} \right.\label{scalingfunctionfull}
\end{equation}

These definitions entirely characterise the spin, directional flaglet transform. The free parameters of the transform are $\lambda,\ \nu,\  N,\ I_0$, and $J_0$. $\lambda$ and $\nu$ define the support of the flaglets in Fourier-Laguerre space and therefore the scales and features the flaglet coefficients extract, as shown in \figref{fig:tiling}.  $I$ and $J$ are the maximum angular and radial flaglet scales captured, respectively, and are fixed by the radial and angular band limits of the signal: \mbox{$I = \lceil \log_\lambda(L-1) \rceil$} and \mbox{$J = \lceil \log_\nu(P-1) \rceil$}, to satisfy \equref{admissibility}. $N$ is used to specify the number of directions to probe. $I_0$ and $J_0$ are the first angular and radial scales of interest (larger scales being captured by the scaling function) and  can be freely chosen provided they satisfy $0\leq I_0 \leq I$ and $0\leq J_0 \leq J$.

The flaglets (and scaling function) tile the angular and radial frequency domains $\ell$ and $p$, as illustrated in \figref{fig:tiling}, while satisfying the admissibility condition of \equref{admissibility}. Wavelets are well localized simultaneously in the spatial domain, both in position and orientation, and the harmonic domain. It is shown in Ref.~\cite{mcewen:2015s2let_localisation} that scale-discretized wavelets on the sphere exhibit excellent concentration properties, both in the scalar setting and in the spin setting. They are also steerable \cite{McEwen:2015so3, McEwen:2013tpa, Wiaux:2007ri}. Flaglets naturally inherit these properties. In particular, as shown in Ref.~\cite{McEwen:2013jpa} for the scalar setting, the flaglet transform forms a tight Parseval frame, \ie the norm of the input signal is conserved.
	
\begin{figure*}
\centering
\includegraphics[height=11cm, trim = 5cm 5.1cm 5.4cm 5.1cm, clip]{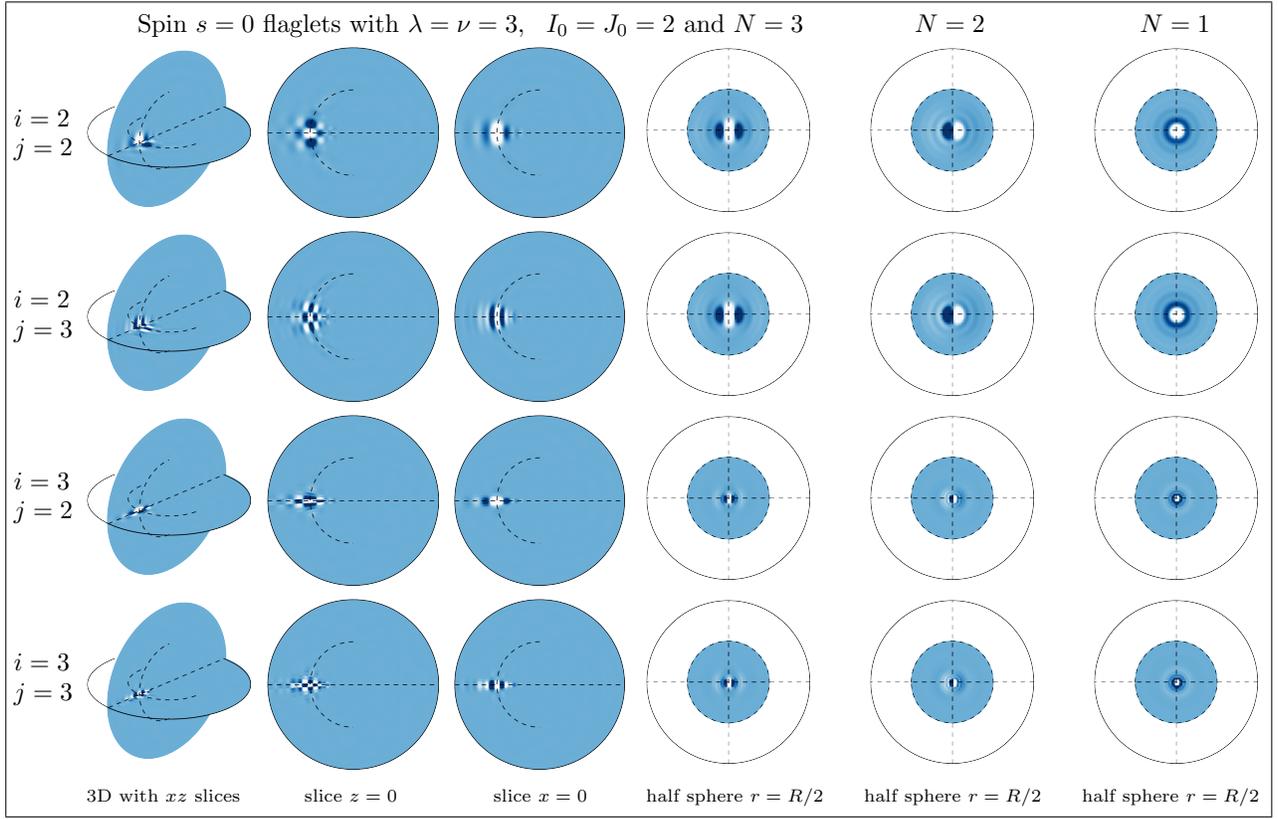}
\caption{Real part of spin-$0$ directional flaglets, plotted up to $r=R/2$ (with $R$ the last node of the radial sampling scheme), showing their excellent localization and directional properties (even and odd for $N=2,3$ respectively). This demonstrates that the flaglet transform can probe radial and angular modes well defined in both real and harmonic space. The radial and angular harmonic modes $\ell$ and $p$ corresponding to the scales $i=j=2,3$ are highlighted as thicker lines in \figref{fig:tiling}. The dashed lines and half sphere show the slices that are plotted in the various subpanels of this figure. Flaglets are localized in both pixel and frequency space, unlike the Fourier-Laguerre basis functions shown in \figref{fig:flagbasis} which are delta functions in harmonic space and therefore not localized in pixel space.}\label{fig:scalarflaglets}
\end{figure*}

\begin{figure*}
\centering
\includegraphics[height=11cm, trim = 5cm 5.1cm 5.4cm 5.1cm, clip]{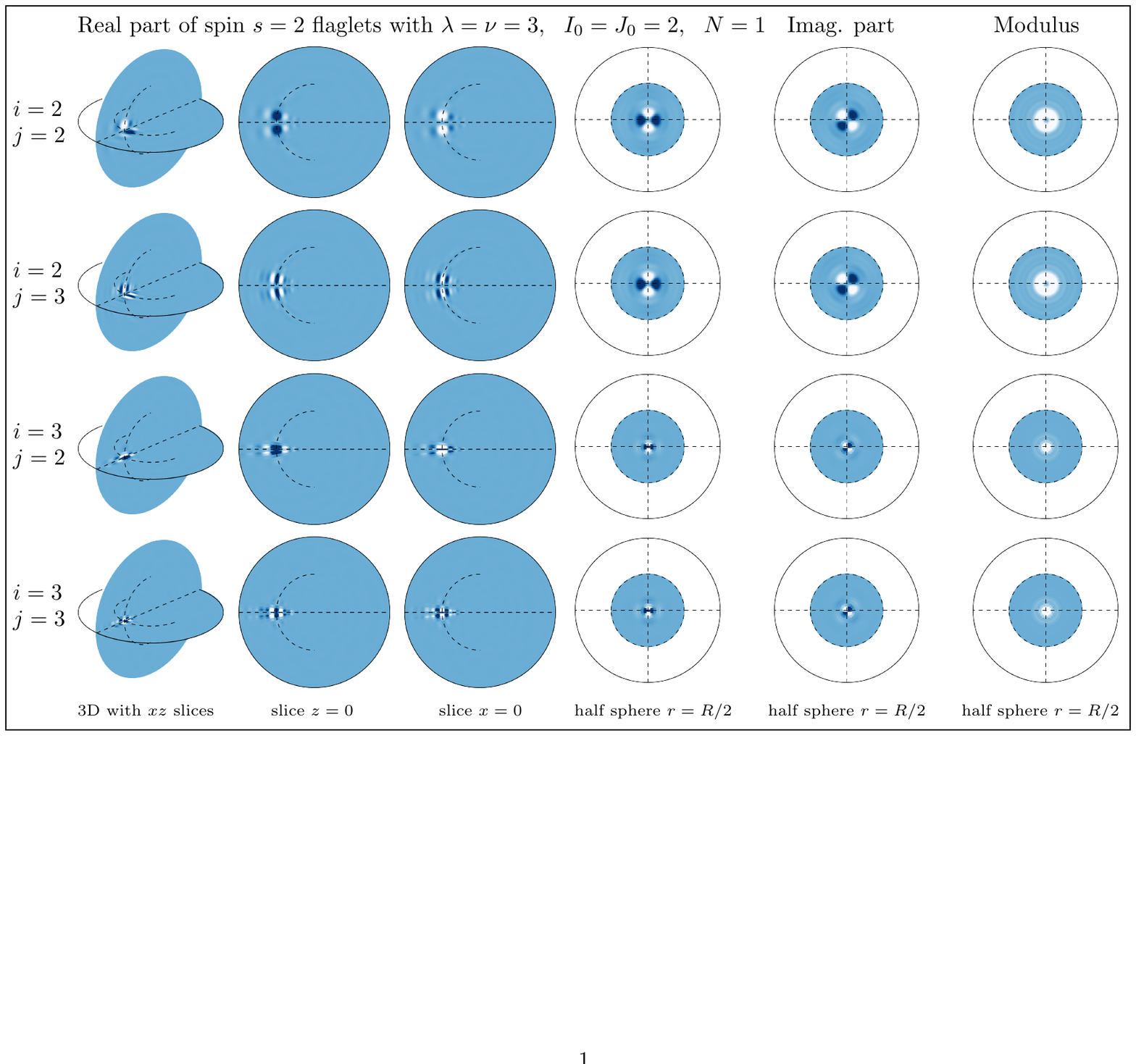}
\caption{Same as \figref{fig:scalarflaglets}, but for spin-$2$ flaglets and showing their real and imaginary parts, as well as the complex modulus. This demonstrates that the flaglets natively support the symmetries of spin signals while being well localized in both real and harmonic space.}\label{fig:spinflaglets}
\end{figure*}

\figrefs{fig:scalarflaglets}{fig:spinflaglets} show spin-$0$ and spin-$2$ flaglets resulting from our construction, with parameters  $\lambda=\nu=3$ and $I_0=J_0=2$. In contrast to the Fourier-Laguerre basis functions shown in \figref{fig:flagbasis}, which are not localized in real space (but are delta functions in harmonic space), all flaglets have good spatial localization properties in both radial and angular dimensions. They are also localized in Fourier-Laguerre space by construction, as shown in \figref{fig:tiling} and \equref{flaglet_lmp}.  Note that the particular $i=2,3$ and $j=2,3$ scales shown in \figrefs{fig:scalarflaglets}{fig:spinflaglets} are highlighted as thicker lines in \figref{fig:tiling}. \figref{fig:scalarflaglets} shows spin-$0$ flaglets (real part only), comparing the $N=2, 3$ flaglets to the axisymmetric $N=1$ case (first shown in Ref.~\cite{Leistedt:2012zx}). The angular components have even and odd symmetry for $N=2, 3$, respectively, by the construction of $\zeta$ \cite{McEwen:2015s2let}. \figref{fig:spinflaglets} shows the real and imaginary parts and the complex modulus of spin-$2$ axisymmetric ($N=1$) flaglets.

%==============================================================================
\subsection{Multiresolution algorithm and implementation}

As described previously, all the integrals and convolutions related to the Fourier-Laguerre and Wigner-Laguerre transforms can be computed exactly via their harmonic space representations and by appealing to sampling theorems to analyze or reconstruct signals in pixel space. Let us consider an input signal ${}_sf\in L^2(\sphere\negsp{1}\times\Rplus)$ with radial and angular band limits $P$ and $L$, respectively. As before, such a signal can be represented exactly by $\sim 2PL^2$ samples and $P L^2$ spherical harmonic coefficients. {The full algorithm to perform the flaglet transform is described by the pipeline below. The first and the third operations are spin Fourier-Laguerre transforms, while the second is the wavelet transform harmonic space filtering operations of \equrefss{eq:wavanalysis_lm}{eq:wavanalysisscal_lm}{eq:wavsynthesis_lm}. 
\begin{eqnarray}
	\nonumber {}_sf(\ang, r) \ \xleftrightarrow[]{\textrm{Fourier-Laguerre}} \
	 {}_sf_{\ell m, p}  
	%\quad\underbrace{\longleftrightarrow}_{\equrefss{eq:wavanalysis_lm}{eq:wavanalysisscal_lm}{eq:wavsynthesis_lm}}\quad
	\ \xleftrightarrow[]{\textrm{Wavelet}} \ \left\{ \left(W_{{}_sf}^{{}_s\Phi}\right)_{\ell m, p}, \left(W_{{}_sf}^{{}_s\Psi^{ij}}\right)^\ell_{m n, p} \right\}
	  \ \xleftrightarrow[]{\textrm{Fourier-Laguerre}}\ % \quad\underbrace{\longleftrightarrow}_{C}\quad
	   \left\{ W_{{}_sf}^{{}_s\Phi}(\ang, r), W_{{}_sf}^{{}_s\Psi^{ij}}(\rhoang, r) \right\}
\end{eqnarray}
}
The flaglet transform will produce flaglet coefficients on SO(3)$\times\Rplus$ for $(I-I_0+1)(J-J_0+1)$ scales, each captured in $\sim 4NPL^2$ samples for an azimuthal band limit $N$ (controlling the number of directions). The additional scaling function is axisymmetric and captured in $\sim 2PL^2$ samples on $\sphere\negsp{1}\times\Rplus$. Therefore, the total number of coefficient scales is $J_{\rm tot} = (I-I_0+1)(J-J_0+1) + 1$. Each one requires the filtering of \equref{eq:wavanalysis_lm}, scaling as $\mathcal{O}(L^2PN)$, and a Wigner-Laguerre transform (or, for the scaling coefficients, filtering of \equref{eq:wavanalysisscal_lm} and a Fourier-Laguerre transform). Therefore the Wigner-Laguerre transforms on SO(3)$\times\Rplus$ will dominate the computation time, and the overall complexity of the flaglet transform is $\mathcal{O}(J_{\rm tot}NPL^3)$. Note that for $N>1$ the scaling function computations are negligible compared to the wavelet computations, in which case we have $J_{\rm tot} = (I-I_0+1)(J-J_0+1)$. 

However, by definition of the flaglets and transform (\equrefs{flaglet_lmp}{eq:wavsynthesis_lm}), the flaglet coefficients have lower band limits than the original signal. Specifically, 
\eqn{
 \left({{}_s\Psi^{ij}}\right)^\ell_{m n, p} \neq 0 \ \ \mathrm{and} \ \left(W_{{}_sf}^{{}_s\Psi^{ij}}\right)^\ell_{m n, p}  \neq \ 0 \
\mathrm{only\ if}  \ \ell\in [\lambda^{i-1}, \lambda^{i+1}], \ m=-\ell,\dots,\ell,  \ n=-N,\dots,N, \ p\in [\nu^{j-1}, \nu^{j+1}].
}
In other words, the flaglet coefficients $W_{{}_sf}^{{}_s\Psi^{ij}}$ have angular and radial band limits $\lambda^{i+1}$ and $\nu^{j+1}$, respectively. Thus, the number of samples needed to capture the information is much smaller than at the full resolution. One can employ a multiresolution algorithm and use the minimum band limits and number of samples for each flaglet coefficient scale (and similarly for the scaling coefficients). This  reduces the complexity of the flaglet transform significantly, to $\mathcal{O}(NPL^3)$, since the computation of the largest scales $ij$ now dominates that of all other scales and the scaling function.

We have implemented the exact and efficient algorithm described previously to compute the spin directional flaglet transform in the existing \textsc{flag} and \textsc{flaglet} codes (\url{http://www.flaglets.org}). The latter relies on the \textsc{ssht} ({\url{http://www.spinsht.org}}), \textsc{s2let} ({\url{http://www.s2let.org}}) and \textsc{so3} (\url{http://www.sothree.org}) codes for the angular transforms and sampling theorems. The \textsc{fftw} ({\url{http://www.fftw.org}}) code is used to compute Fourier transforms. The core algorithms of the Fourier-Laguerre, Wigner-Laguerre and flaglet transforms are implemented in \textsc{C} and are able to handle large band limits and billions of samples on $\sphere\negsp{1}\times\Rplus$ and SO(3)$\times\Rplus$. Interfaces to the core functions as well as convenient data manipulation and plotting routines are provided in \textsc{Matlab} and \textsc{Python}.

%==============================================================================
%==============================================================================
\section{Application to 3D weak lensing}\label{sec:weaklensingapplication}

In previous sections we developed the spin flaglet formalism to extract content localized in both space and frequency from 3D signals of arbitrary spin, which has the added benefit that radial and angular modes are separable.  This is a general framework that can be applied to arbitrary 3D spin signals. 
We now apply this approach to represent 3D cosmic shear and other weak lensing observables, and relate the covariance of their flaglet decompositions to the lensing power spectrum. In particular, we highlight how one can exploit the properties of flaglets (separability and simultaneous localization in frequency and pixel space) to deal with partial sky coverage and small-scale modeling uncertainties.  We also examine the approximations commonly adopted in cosmic shear analyzes and apply them to the flaglet covariance of weak lensing observables. 
{Note that we focus on using axisymmetric wavelets to increase simplicity and readability of the final estimator. However, the latter is straightforwardly extended to directional flaglets, and this does not affect any of the advantages and properties discussed below.}

%==============================================================================
\subsection{3D cosmic shear}

We relate the covariance of the flaglet representation of the cosmic shear field to the lensing power spectrum. 
As shown in \secref{sec:weaklensingintro}, the 3D cosmic shear signal ${}_2\gamma$ is characterised by the lensing power spectrum $C^{\phi\phi}_\ell(k,\prim{k})$ defined in Fourier-Bessel space by
\eqn{
	\langle \ {}_2\gamma_{\ell m}(k) \ {}_2\gamma^*_{\prim{\ell}\prim{m}}(\prim{k}) \ \rangle \ = \ \frac{1}{4} (N_{\ell, 2})^2 C^{\phi\phi}_\ell(k,\prim{k}) \delta^K_{\ell \prim{\ell}}\delta^K_{m \prim{m}},
}
where all relevant physical effects are assumed to be modelled (as discussed in \secref{sec:weak_lensing:cosmology}). As in \secref{sec:weaklensingintro}, the brackets in this section denote ensemble average over cosmological realisations, rather than inner products (which are distinguished by a vertical bar between the functions of the inner product).

 We now consider the wavelet transform of ${}_2\gamma$, using axisymmetric spin flaglets (\ie $N=1$). In this case all directional convolutions $\convdir$ become axisymmetric convolutions $\convaxisym$.  The flaglet coefficients live on $\sphere\times\Rplus$ and read
\equ{
W_{{}_2\gamma}^{{}_2\Psi^{ij}}(\ang,r)  = ({}_2\gamma \convaxisym {}_2\Psi^{ij})(\ang,r).
}
We consider the covariance between flaglet scales,
\eqn{
	C^{ij,\prim{i}\prim{j}}(\ang,\prim{\ang},r,\prim{r}) = \langle \ W_{{}_2\gamma}^{{}_2\Psi^{ij}}(\ang,r) \ W_{{}_2\gamma}^{{}_2\Psi^{\prim{i}\prim{j}}*}(\prim{\ang},\prim{r}) \ \rangle.
}
Injecting the Fourier-Bessel decomposition of ${}_2\gamma$ into the integral form of the convolution of $W_{{}_2\gamma}^{{}_2\Psi^{ij}}(\ang,r)$, one obtains
\eqn{
&&C^{ij,\prim{i}\prim{j}}(\ang,\prim{\ang},r,\prim{r}) \ = \
	 \frac{2}{\pi} \sum_{\ell m} \frac{(N_{\ell, 2})^2}{4}  \int_\Rplus\negsp{2} \d k k^2\negsp{1} \int_\Rplus\negsp{2} \d \prim{k} k^{\prime 2} 
	C^{\phi\phi}_\ell(k,\prim{k}) \ {}_2F_{\ell m}^{ij}(k,\ang,r) \ {}_2F_{\ell m}^{\prim{i}\prim{j}}{}^*(\prim{k},\prim{\ang},\prim{r}),
}
where ${}_2F_{\ell m}^{ij}(k,\ang,r)$ is the wavelet transform of the Fourier-Bessel basis functions:
\eqn{
  {}_2F_{\ell m}^{ij}(k,\ang,r) \ &=& \ ({}_2Y_{\ell m} j_\ell(k\cdot) \convaxisym {}_2\Psi^{ij})(\ang,r) \\
 &=&  \int_{\sphere}\d\Omega(\prim{\ang})  \int_{\Rplus} \negsp{2} \d \prim{r}r^{\prime2} {}_2Y_{\ell m}(\prim{\ang}) j_\ell(k\prim{r}) \ ( \mathcal{R}_{\ang} \mathcal{T}_r \ {}_2\Psi^{ij} )^*(\prim{\ang}, \prim{r}) .
}
Note that this formalism can be straightforwardly extended to directional wavelets: directional convolutions then replace axisymmetric convolutions, and the angles $\ang,\prim{\ang}\in\sphere$ become $\rhoang,\prim{\rhoang}\in$ SO(3) since the flaglet coefficients then live on SO(3)$\times\Rplus$.

The previous expression can be simplified by appealing to the harmonic representation of the translation and rotation operators, which when applied to axisymmetric flaglets ${}_2\Psi^{ij}$ (\eg Refs.~\cite{Leistedt:2012gk, McEwen:2015s2let}) yield
\eqn{
	(\mathcal{T}_r \mathcal{R}_\ang \ {}_2\Psi^{ij})_{\ell m, p} = \sqrt{\frac{4\pi}{2\ell+1}} K_p^*(r) Y_{\ell m}^*(\ang) \ {}_2\Psi^{ij}_{\ell 0, p}.
}
The functions ${}_2F_{\ell m}^{ij}(k,\ang,r)$ then take the simple form 
\eqn{
	&&{}_2F_{\ell m}^{ij}(k,\ang,r) \ = \ \sqrt{\frac{4\pi}{2\ell+1}} Y_{\ell m}^*(\ang) {\sum_p \mathcal{J}_{\ell, p}(k)\  K_p(r)  \ {}_2\Psi^{ij*}_{\ell 0, p}} 
  \ = \ \sqrt{\frac{4\pi}{2\ell+1}} Y_{\ell m}^*(\ang) \ {}_2\mathcal{H}_{\ell}^{ij}(k,r), \label{eq:Ffunctions}
 }
where
\eqn{  
 {}_2 \mathcal{H}_{\ell}^{ij}(k,r) = \sum_p \mathcal{J}_{\ell, p}(k)\  K_p(r)  \ {}_2\Psi^{ij*}_{\ell 0, p}.
}
As shown previously, the functions $\mathcal{J}_{\ell, p}$ admit an analytical form \cite{Leistedt:2012zx} and can be tabulated. Furthermore, ${}_2\mathcal{H}_{\ell}^{ij}(k,r)$, and consequently ${}_2F_{\ell m}^{ij}(k,\ang,r)$, is straightforward to compute since only a one dimensional summation over a finite number of $p$ samples is required.

Further simplifications can be made by exploiting the spherical harmonic addition theorem:
\equ{
  \sum_m Y_{\ell m}(\ang)Y^*_{\ell m}(\prim{\ang}) = \frac{2\ell+1}{4\pi} P_\ell(\ang \cdot \prim{\ang}), \label{eq:sumYlm}
  }
where $P_\ell$ are the Legendre polynomials.  By the additional theorem one can infer that the wavelet covariance only depends on the angle $\Delta\theta$, where $\ang \cdot \prim{\ang} = \cos(\Delta\theta)$. Thus, $C^{ij,\prim{i}\prim{j}}(\ang,\prim{\ang},r,\prim{r})$ becomes
\eqn{
	C^{ij,\prim{i}\prim{j}}(\Delta\theta,r,\prim{r})    \ = \
	 \frac{2}{\pi} \sum_{\ell} \frac{(N_{\ell, 2})^2}{4}  \int_\Rplus\negsp{2} \d k k^2\negsp{1} \int_\Rplus\negsp{2} \d \prim{k} k^{\prime 2} \
	C^{\phi\phi}_\ell(k,\prim{k}) \ P_\ell(\Delta\theta) \ {}_2\mathcal{H}_{\ell}^{ij}(k,r) \ {}_2\mathcal{H}_{\ell}^{\prim{i}\prim{j}}{}^*(\prim{k},\prim{r}) .\label{flagletcov}
}
The angular dependence of the flaglet covariance depends only on the angular separation between pixels on the sky. This is expected since we are analyzing the 2-point fluctuations of the signal with axisymmetric flaglets and have assumed statistical homogeneity and isotropy. 

\equref{flagletcov} is one of the key results of this paper. One can build a full likelihood analysis of 3D weak lensing observables and constrain cosmological models and parameters using spin flaglets (similar to that of Ref.~\cite{Kitching:2014dtq}). On the left hand side of \equref{flagletcov} is a quantity that can be computed from data using a wavelet decomposition ({here using axisymmetric flaglets, but a similar expression can be obtained with directional flaglets).} On the right hand side is the theoretical expectation of this quantity related to cosmological parameters through the 3D lensing power spectrum $C^{\phi\phi}_\ell(k,\prim{k})$. {Even though $C^{\phi\phi}_\ell(k,\prim{k})$ mixes radial and angular modes, these are isolated into scales $(i,j,i^\prime,j^\prime)$ by the kernels ${}_2\mathcal{H}_{\ell}^{ij}(k,r)$.} In practice, the flaglet covariance can be calculated for a pair of flaglet coefficients $ij$ and $\prim{i}\prim{j}$ and a pair of radii $r$ and $\prim{r}$ by averaging over pairs of pixels separated by $\Delta \theta$. This approach naturally accounts for the 2+1D nature of the shear observables and takes advantage of the convenient connection between the Fourier-Laguerre and Fourier-Bessel transforms. {It takes full advantage of the separability of the Fourier-Laguerre and flaglet transforms.} Going through flaglet space allows one to exploit the excellent localization properties of flaglets in both pixel and frequency space, a property absent in all alternative approaches.  Simultaneous localization in both pixel and frequency space allows one to cut or filter regions of the sky due to unobserved, unreliable or contaminated data, at the same time as cutting or filtering harmonic modes.  For example, one may wish to cut or filter small-scale harmonic modes where physical modeling is less certain, \eg high-$k$ (and $k^\prime$) modes from $C^{\phi\phi}_\ell(k,\prim{k})$. This is possible by studying the kernels ${}_2\mathcal{H}_{\ell}^{ij}(k,r)$ and only considering the flaglet scales $ij,i^\prime j^\prime$ that probe the physical scales of interest in $C^{\phi\phi}_\ell(k,\prim{k})$. Thus, the spin flaglet 3D weak lensing approach is a flexible way to avoid the complications of correlation functions and Fourier-Bessel representations, which are typical of standard approaches.

%==============================================================================
\subsection{Generalization to other weak lensing quantities}
\label{flexion}
As shown in \equref{chidef}, spin-2 shear distortions are not the only effect of matter fluctuations on the observed properties of galaxies \cite{Bacon:2008zj}. Other observables are expected to be measured at much lower signal-to-noise ratio than the shear distortion but they nevertheless contain important cosmological information. Thus, we generalize the flaglet covariance derived previously to other lensing distortions of different spin.  For this purpose we introduce a general scaling factor
\eqn{
	M_{\ell, s} &=& \left\{ \begin{array}{lll}  
		-\frac{1}{2} \sqrt{\ell^2(\ell+1)^2}, \quad\quad &{\rm if} \ s = 0 \vspace*{2mm}\\
	\frac{1}{6} [ \sqrt{\ell(\ell+1)(\ell-1)^2(\ell+2)^2} + 2\sqrt{\ell^3(\ell+1)^3} ], \quad\quad &{\rm if} \ s = 1 \vspace*{2mm}\\
	 \frac{1}{2} \sqrt{\frac{(\ell+2)!}{(\ell-2)!}} \ =\ \frac{1}{2} N_{\ell, 2}, \quad\quad &{\rm if} \ s = 2 \vspace*{2mm}\\
	 -\frac{1}{2} \sqrt{\frac{(\ell+3)!}{(\ell-3)!}} \ =\  -\frac{1}{2} N_{\ell, 3}, \quad\quad &{\rm if} \ s = 3  \end{array}\right.,
}
so that we can write ${}_s\chi_{\ell m}(k) = M_{\ell, s} \phi_{\ell m}(k)$, with ${}_s\chi$ defined in \equref{chidef}. In this unified approach, we have
\eqn{
	&&C^{ij,\prim{i}\prim{j}}(\ang,\prim{\ang},r,\prim{r}) \ = \
	 \frac{2}{\pi}  \sum_{\ell m} (M_{\ell, s})^2  \int_\Rplus\negsp{2} \d k k^2\negsp{1} \int_\Rplus\negsp{2} \d \prim{k} k^{\prime 2} 
	C^{\phi\phi}_\ell(k,\prim{k}) \ {}_sF_{\ell m}^{ij}(k,\ang,r) \ {}_sF_{\ell m}^{\prim{i}\prim{j}}{}^*(\prim{k},\prim{\ang},\prim{r})
}
with ${}_sF_{\ell m}^{ij}(k,\ang,r)$ generalizing \equref{eq:Ffunctions} to other values of spin. As before, the extension to directional wavelets is straightforward. Exploiting the spherical harmonic addition property, again, leads to the simplification
\eqn{
\label{key}
	C^{ij,\prim{i}\prim{j}}(\Delta\theta,r,\prim{r})    \ = \
	 \frac{2}{\pi} \sum_{\ell} {(M_{\ell, s})^2}  \int_\Rplus\negsp{2} \d k k^2\negsp{1} \int_\Rplus\negsp{2} \d \prim{k} k^{\prime 2} 
	\ C^{\phi\phi}_\ell(k,\prim{k}) \ P_\ell(\Delta\theta) \ {}_s\mathcal{H}_{\ell}^{ij}(k,r) \ {}_s\mathcal{H}_{\ell}^{\prim{i}\prim{j}}{}^*(\prim{k},\prim{r}) .
}
with ${}_s{\mathcal{H}_{\ell}^{ij}(k,r)}$ defined as previously. 
This expression generalizes the 3D cosmic shear power spectrum to the case of  lensing distortions of other spin. As before, this quantity can be calculated from data by averaging over pairs of pixels separated by $\Delta\theta$ in bins of $ij,\  \prim{i}\prim{j},\ r,\prim{r}$. The theoretical prediction through the lensing potential, the Legendre polynomials and the ${}_s\mathcal{H}$ kernels is readily computable. {Finally, this formalism can also be extended to support cross-terms between the various spin components, which would be required for a combined analysis.}
 
%==============================================================================
\subsection{Approximations}

We have related the covariance of weak lensing observables in flaglet space to the underlying 3D lensing power spectrum in the general setting. Several approximations are used in the literature to make the computation of shear and other lensing quantities more approachable. Some of these are lossless---or nearly so---while others result in a loss of information with respect to the general case. The most commonly used approximation is to use cosmic shear `tomography', in which flat-sky and Limber approximations are typically also made.  We apply these approximations below to the flaglet covariance of spin lensing quantities, where appropriate, and show that they result in representations that are more readily computable but are a lossy representation of the 3D lensing power spectrum.

%==============================================================================
\subsubsection{Flat-sky approximation} 

Weak gravitational lensing of galaxy images has the strongest signal-to-noise on the scales of groups or clusters (\eg Ref.~\cite{Bartelmann:1999yn}) and therefore there is relatively little signal in the power spectrum on large angular scales. In addition, for surveys with relatively small sky coverage and/or surveys with several small observational fields, effects due to the spherical geometry of the setting are expected to have a relatively small impact.  However, for future surveys (\eg Euclid \cite{Amendola:2012ys, Laureijs:2011gra} and LSST \cite{Abell:2009aa}) with observations over increasingly greater coverages of the sky, flat-sky approximations will become increasingly less accurate.

The flat-sky approximation assumes that the angular extent of the observational field is small and hence the geometry of the angular component is assumed to be planar (\ie Euclidean).  In this setting the spherical harmonics are approximated by the product of Bessel functions (of the first kind) and complex exponentials \cite{Castro:2005bg, Hu:2000ee}.  In the case of \equrefs{flagletcov}{key}, however, the sum over the product of spherical harmonic functions reduces to a Legendre polynomial (due to the spherical harmonic addition theorem), which in the flat-sky case would be approximated by a zeroth order Bessel function.  However, given the computational ease of computing Legendre polynomials the full-sky case is readily computable and so a flat-sky approximation is not required. This differs to the computation of $C^{\phi\phi}_\ell(k,\prim{k})$, where a flat-sky approximation can reduce computational time considerably and has so-far been assumed in current applications to existing data \cite{Kitching:2006mq, Kitching:2014dtq}, where sky coverage is relatively small.

%==============================================================================
\subsubsection{Limber approximation} 

The Limber approximation \cite{LoVerde:2008re} assumes that the evolution of radial modes is small over the survey volume under consideration.  Under this approximation the spherical Bessel functions may be approximated by Dirac delta functions:
\equ{
  k^{1/2}j_{\ell}(kr)
  \simeq
  \sqrt{\frac{\pi}{2\ell+1}}\delta^{\rm D}(\ell+1/2 - kr),
  \label{eq:limber2}
}
as shown in Refs.~\cite{Munshi:2010ny, Kitching:2014lga}, and rederived in \appref{app:limber} for the spherical Bessel convention adopted herein.

The use of this approximation simplifies matters considerably: an 8-nested integral equation for $C^{\phi\phi}_\ell(k,\prim{k})$ becomes a single integral (see, \eg Refs.~\cite{Kitching:2010wa, Kitching:2014dtq}). {However, even though this approximation primarily affects large scales $\ell \lesssim 100$, it is severe and can cause errors in the inferred cosmological parameters of tens of percent \cite{Kitching:2010wa}. }
% This mapping can be placed into any integral or sum in which the Bessel function appears. In fact, this does not explicitly convert a 3D wavevector into a 2D wavevector, i.e. $k =(k_x, k_y, k_z) \rightarrow (k_x, k_y)$, which is what the original cosmological application the Limber approximation does in Ref.~\cite{Kaiser:1996tp}, but this transformation is implicit when an integral over the delta function is performed. 
Exploiting the Limber approximation simplifies the $\mathcal{J}_{\ell, p}(k)$ functions to 
\equ{
  \mathcal{J}_{\ell, p}(k) 
  \simeq 
  \sqrt{\frac{\pi}{2}} \ \frac{(\ell+1/2)^{3/2}}{k^{7/2}} \ K^*_p\Bigl(\frac{\ell+1/2}{k}\Bigr),
  \label{eq:besselflagcoefs_limber}
}
and the ${}_2\mathcal{H}^{ij}_{\ell}(k,r)$ kernels to
\equ{
  {}_2\mathcal{H}^{ij}_{\ell}(k,r) 
  \simeq 
  \sqrt{\frac{\pi}{2}} \ \frac{(\ell+1/2)^{3/2}}{k^{7/2}} \ 
  \sum_p K^*_p\Bigl(\frac{\ell+1/2}{k}\Bigr)
  K_p(r)  \ {}_2\Psi^{ij}_{\ell 0, p}{}^*.
  \label{eq:gcal_limber}
}
This is a significant simplification since, although $\mathcal{J}_{\ell, p}(k)$ already does not need to be computed by the integration of products of Bessel and Laguerre functions (see \equref{besselflagcoefs}) and can be computed analytically (as shown in Ref.~\cite{Leistedt:2012zx}), the analytical computation can also be challenging for high-$\ell$.  \equref{eq:besselflagcoefs_limber} and, consequently, \equref{eq:gcal_limber} can be readily computed both accurately and efficiently for all $\ell$.
 % Similar transformation can be made in the $C^{\phi\phi}_\ell(k,\prim{k})$ function, but note that it is still a function of $\ell$ and $(k$, $\prim{k})$.

%==============================================================================
\subsubsection{Tomographic approximation} 

The tomographic approximation is a combination of the the flat-sky and Limber approximation with the addition of a discretisation of the signal along the radial direction into a series of redshift bins.  Within each redshift bin the projected 3D contribution, from a range of galaxies with redshifts assigned to that bin, is used to construct a 2D power spectrum. This binning in redshift is lossy as any evolution that occurs in the signal on scales smaller than the bin widths is lost.  Nevertheless, due to computational and conceptual ease it is the most widely used approach. 

To relate the wavelet decomposition of the data to the tomographic power spectrum we do not require radial transforms but can simply relate the wavelet covariance to the cosmic shear tomography power spectrum using the approximation 
\equ{
  C^{\phi\phi}_\ell(k,\prim{k}) \simeq C^{\phi\phi}_\ell(r,\prim{r})\delta^D(\ell+1/2 - kr)\delta^D(\ell+1/2 - \prim{k}\prim{r}).
}
By this tomographic approximation, \equref{key} reduces to
\eqn{
	C^{ij,\prim{i}\prim{j}}(\Delta\theta,r,\prim{r})    \ \simeq \
	 \frac{2}{\pi} \sum_{\ell} {(M_{\ell, s})^2} \ 
         \frac{(\ell+1/2)^4}{r^3 \prim{r}{}^3} \ 
	C^{\phi\phi}_\ell(r,\prim{r}) \ P_\ell(\Delta \theta) \ 
        {}_2\mathcal{H}_{\ell}^{ij}\Bigl(\frac{\ell+1/2}{r},r\Bigr) \ 
        {}_2\mathcal{H}_{\ell}^{\prim{i}\prim{j}}{}^*\Bigl(\frac{\ell+1/2}{\prim{r}},\prim{r}\Bigr) ,
}
where the $\mathcal{H}^{ij}_{\ell}$ kernels are those given in the Limber-approximated case of \equref{eq:gcal_limber}.  The final 
approximation for tomography with binning in redshift is 
\equ{
  C^{ijj'k'}(\Delta\theta, r_{{\rm bin}}(z_{a}),r_{{\rm bin}}(z_{\prim{a}}))
  \simeq 
  \int_\Rplus {\rm d}r \int_\Rplus {\rm d}\prim{r} \ 
  C^{ij,\prim{i}\prim{j}}(\Delta\theta,r,\prim{r}) \
  W(r,\prim{r},r_{{\rm bin}}(z_{a}),r_{{\rm bin}}(z_{\prim{a}})),
}
where the weight function $W$ describes if a galaxy is in the bin-pair combination 
$(r_{{\rm bin}}(z_a),r_{{\rm bin}}(z_{\prim{a}}) )$ or not.  Bin weight functions are 
usually designed as top-hat functions in spectroscopic 
redshift $z_{a}$ and corrections are applied for `leakage' due to photometric redshifts (see, \eg Ref.~\cite{Kitching:2010wa}). The methods presented here can also be extended to deal with redshift uncertainties and more optimal weight functions.

% The same integral is  then performed on on the left-hand side of the equation resulting in a `tomographic' cosmic shear power 
% spectrum $C^{\phi\phi}_{\ell}(z_{\alpha},z_{\beta})$.

% Beyond these one can also make even further approximations that involve the angle transform from Fourier space to 
% real space to create correlation function measurements. 

%==============================================================================
\section{Conclusions}\label{sec:conclusion}

We have developed a novel approach to analyze weak gravitational lensing observations using 3D wavelets (flaglets) and the Fourier-Laguerre transform. We have shown that the covariance of the flaglet coefficients of weak lensing observables (such as cosmic shear, size distortions and flexion) can be directly related to the 3D lensing power spectra, exploiting the analytical connection between the Fourier-Bessel and Fourier-Laguerre transforms. Thanks to the separability of their radial and angular components, the Fourier-Laguerre and flaglet transforms can take advantage of existing fast algorithms and sampling theorems developed for the radial line, the sphere, and the rotation group. Furthermore, this approach is suited to dealing with the 2+1D nature of weak lensing observations and cosmological data sets in general. For example, typical data complications and systematic uncertainties are given on the sky and in the radial direction separately. The excellent localization properties of flaglets in both pixel and frequency space can be exploited to deal with these effects and alleviate the complications of weak lensing analyzes (\eg complicated sky masks and uncertainties of the small-scale modeling) without complicating the estimation of the flaglet coefficients and covariance. Therefore, the method introduced here is a powerful alternative to common real and harmonic-space approaches where these effects are difficult to deal with \cite{Kitching:2006mq, Kitching:2014dtq}. Lensing observables and covariances are more easily modelled in the Fourier-Bessel basis, where unreliable or unmodelled small scales can also be filtered out. However, the mode-mixing due to partial sky coverage is difficult to handle in Fourier-Bessel space. By contrast, real space correlation function measurements can naturally deal with complicated survey window functions, but filtering scales and modeling covariances is then significantly more difficult. The dual localization of wavelets in pixel and harmonic space gives them the advantages of both approaches. In future work we will demonstrate the ability of this new approach to robustly extract cosmological information by applying it to simulations and real observations of weak lensing signals.
 
We have updated the publicly available \textsc{flag} and \textsc{flaglet} (\url{http://www.flaglets.org}) codes to compute the spin Fourier-Laguerre and flaglet transforms. These rely on the fast algorithms implemented in the following codes, also publicly available: \textsc{s2let} ({\url{http://www.s2let.org}}) for the spin directional spherical wavelets, \textsc{ssht} ({\url{http://www.spinsht.org}}) for the spherical harmonics transform, \textsc{so3} (\url{http://www.sothree.org}) for the Wigner transform, and \textsc{fftw} (\url{http://www.fftw.org}) for the Fourier transform. In all these codes, the core routines are implemented in \textsc{C}, and we have provided numerous wrappers and interfaces in \textsc{Python}, \textsc{Idl}, and \textsc{Matlab}, to call the core routines, and to manipulate and visualize data sets. As highlighted in Ref.~\cite{Leistedt:2012zx} for the scalar setting, these codes can deal with large band limits and large 2D and 3D data sets, which can be pixelized into billions of samples and manipulated with no loss of information. 

\acknowledgments

B.~Leistedt was supported by the Impact and Perren funds. B.~Leistedt and H.~V.~Peiris are supported by the European Research Council under the European Community's Seventh Framework Programme (FP7/2007-2013) / ERC grant agreement no 306478-CosmicDawn. J.~D.~McEwen is supported by the Engineering and Physical Sciences Research Council (grant number EP/M011852/1).

%==============================================================================
\appendix

%==============================================================================
\section{Properties of special functions}\label{app:spinsha}

We concisely review the orthogonality and completeness properties of the special functions considered throughout this paper, of which we make continual use.  In particular, we consider the spherical Bessel functions, the Laguerre polynomials, the spin spherical harmonics, for which we also review the spin raising and lowering operators, and finally the Wigner functions.

The spherical Bessel function (of the first kind) of order $\ell$, denoted by $j_\ell$, is defined by
\eqn{
  j_\ell(x) = \sqrt{\frac{\pi}{2x}} J_{\ell+1/2}(x),
}
where $J_\ell$ is the standard Bessel function of the first kind. The closure relation for the spherical Bessel functions is given by
\equ{
  \int_\Rplus {\rm d}r r^2 j_\ell(kr) j_\ell(\prim{k}r) 
  = \frac{\pi}{2 k^2} \delta^{\rm D}(k - \prim{k}),
}
which plays the role of both completeness and orthogonality relations in the continuous setting (note that the spherical Bessel transform does not admit a 
sampling theorem leading to exact quadrature \cite{McEwen:2013jpa, Leistedt:2012zx}). 
% The forward and inverse spherical Bessel transforms of a square integrable complex signal on the radial line $f\in L^2(\Rplus)$ are given by 
% \begin{eqnarray}
%   f_\ell(k) &=& \sqrt{\frac{2}{\pi}}  \int_{\Rplus} \negsp{2}  \d r r^2  f(r)   j^*_\ell(kr) \\
%   f(r) &=&  \sqrt{\frac{2}{\pi}} \int_\Rplus\negsp{2} \d k k^2  f_\ell(k)   j_\ell(kr) .
% \end{eqnarray}
The spherical Laguerre basis functions $K_p$ are related to the standard Laguerre polynomials by \equref{eq:laguerrebasis}.  The orthogonality of the spherical Laguerre functions reads
\eqn{
   \langle K_p | K_q \rangle = \int_{\mathbb{R}^+} {\rm d} r r^2 K_p(r) K^*_q(r) = \delta^{\rm K}_{pq}, 
}
while completeness is inherited from the completeness of polynomials \cite{Leistedt:2012zx}.  Both the spherical Bessel and Laguerre functions form a complete basis for the representation of functions defined on the radial line $\Rplus$.

The orthogonality and completeness of the spin spherical harmonic functions ${}_sY_{\ell m}$ read
\eqn{
  \langle {}_sY_{\ell m} | {}_sY_{\prim{\ell} \prim{m}} \rangle = 
  \int_{\sphere}  \d\Omega(\ang)  
  {}_sY_{\ell m}(\ang) {}_sY_{\prim{\ell} \prim{m}}^*(\ang)
  = \delta^{\rm K}_{\ell \prim{\ell}} \delta^{\rm K}_{m \prim{m}}
}
and
\eqn{
  \sum_{\ell m} {}_sY_{\ell m}(\ang) {}_sY_{\ell m}^*(\prim{\ang})
  = \delta^{\rm D}(\ang - \prim{\ang}) 
  = \delta^{\rm D}(\cos\theta - \cos\prim{\theta}) 
  \delta^{\rm D}(\phi - \prim{\phi}) ,
}
respectively.  The spin spherical harmonics are the canonical (complete) basis for the representation of functions defined on the sphere $\sphere$, while accounting for the symmetry of spin-$s$ signals.
Spin raising and lowering operators, \mbox{$\eth$ and $\bar{\eth}$} respectively, increment and decrement the spin order of a spin-$s$ function and read \cite{Goldberg:1966uu, Newman:1966ub, Zaldarriaga:1996xe, Kamionkowski:1996ks} 
\eqn{
  \eth &=& - \sin^s \theta\left( \frac{\partial}{\partial\theta} + \frac{\im}{\sin \theta} \frac{\partial}{\partial\phi} \right) \sin^{-s}\theta,\\
  \bar{\eth} &=& - \sin^{-s} \theta\left( \frac{\partial}{\partial\theta} - \frac{\im}{\sin \theta} \frac{\partial}{\partial\phi} \right) \sin^{s}\theta ,
}
respectively. Using these operators, spin-$s$ spherical harmonics ${}_sY_{\ell m}$  can be constructed from scalar (spin-$0$) harmonics by
\eqn{
  {}_sY_{\ell m}(\ang) &=& N_{\ell, -s} \ \eth^s Y_{\ell m}(\ang), \quad \ {\rm for}\  0\leq s\leq \ell\\
  {}_sY_{\ell m}(\ang) &=& N_{\ell, +s} \ \bar{\eth}^{-s} Y_{\ell m}(\ang), \quad {\rm for}\ -\ell\leq s\leq 0. \quad\nonumber
}
We assumed the Condon-Shortley phase convention and define the factor
\eqn{
  N_{\ell, s} = \sqrt{ \frac{(\ell+s)!}{(\ell-s)!}}.
}

The orthogonality and completeness of the Wigner functions $D^{\ell}_{m n}$ read
\eqn{
  \langle D^{\ell}_{m n} | D^{\prim{\ell}}_{\prim{m} \prim{n}} \rangle = 
  \int_{\rm SO(3)} \d\mu(\rhoang) 
  D^{\ell}_{m n}(\rhoang)
  D^{\prim{\ell}*}_{\prim{m} \prim{n}}(\rhoang)  
  = \frac{8\pi}{2\ell+1}
  \delta^{\rm K}_{\ell \prim{\ell}} \delta^{\rm K}_{m \prim{m}}
  \delta^{\rm K}_{n \prim{n}}
}
and
\eqn{
  \sum_{\ell m n} D^{\ell}_{m n}(\rhoang) D^{\ell *}_{m n}(\prim{\rhoang})
  = \delta^{\rm D}(\rhoang - \prim{\rhoang}) 
  = 
  \delta^{\rm D}(\alpha - \prim{\alpha}) 
  \delta^{\rm D}(\cos\beta - \cos\prim{\beta}) 
  \delta^{\rm D}(\gamma - \prim{\gamma}),
}
respectively.  The Wigner functions provide a complete basis for the representation of functions defined on the rotation group SO(3). Note that we adopt $D^{\ell *}_{m n}$ as basis functions since this simplifies connections to wavelet transforms.

%==============================================================================
\section{Limber approximation}\label{app:limber}

The extended Limber approximation is derived in Ref.~\cite{LoVerde:2008re}.  By comparing Eqs.~(5) and (12) of Ref.~\cite{LoVerde:2008re} and neglecting higher order terms, we find:
\equ{
  \int_\Rplus {\rm d}k k^2 j_{\ell}(kr) j_{\ell}(k\prim{r}) F(k)
  \simeq \frac{\pi}{2 r^2} F\Bigl(\frac{\ell+1/2}{r}\Bigr) \delta^{\rm D}(r - \prim{r}),
  \label{eq:limber1}
}
which was noted previously by \cite{Munshi:2010ny}.  It follows that the Limber approximation can be represented by the following approximation of the spherical Bessel functions:
\equ{
  k^{1/2}j_{\ell}(kr)
  \simeq
  \sqrt{\frac{\pi}{2\ell+1}}\delta^{\rm D}(\ell+1/2 - kr).
}
We note a difference of a factor of $k^{1/2}$ compared to \cite{Munshi:2010ny} and an additive factor of $1/2$ in the argument of $F$ in \equref{eq:limber1}.  Note that \cite{Munshi:2010ny} consider a different convention for the spherical Bessel transform to that considered herein (which differ by a factor of $k$).  We follow the same convention as \cite{Heavens:2003jx}.

%==============================================================================

\bibliography{bib}

\end{document}